\documentclass[aps,preprint,amsmath,amssymb]{revtex4}
\usepackage{graphicx,color}
\usepackage{subfigure}
\def\be{\begin{eqnarray}}
\def\ed{\end{eqnarray}}
\def\non{\nonumber}

\def\Hd{H^\dagger}
\def\Pd{\Phi^\dagger}
\def\Dd{\Delta^\dagger}

\begin{document}


\title{  Inert  Dark Matter in Type-II Seesaw}

\author{ \bf Chuan-Hung Chen$^{a}$\footnote{Email:
physchen@mail.ncku.edu.tw} and Takaaki Nomura$^{a}$\footnote{Email: numura@mail.ncku.edu.tw} }

\affiliation{ $^{a}$Department of Physics, National Cheng-Kung
University, Tainan 701, Taiwan  }

\date{\today}

\begin{abstract}

Weakly interacting massive particle (WIMP) as a dark matter (DM) candidate  is further inspired by  recent AMS-02 data, which  confirm  the excess of positron fraction observed earlier by PAMELA and Fermi-LAT experiments. Additionally, the excess of positron+electron flux is still significant in the measurement of Fermi-LAT. 
For solving the  problems of massive neutrinos and observed excess of cosmic-ray, we study the model with an inert Higgs doublet (IHD) in the framework of type-II seesaw  model by imposing a $Z_2$ symmetry on the IHD, where the lightest particle of IHD is the DM candidate and the neutrino masses originate from  the Yukawa couplings of Higgs triplet and leptons. We calculate the cosmic-ray production in our model by using  three kinds of neutrino mass spectra, which are classified by normal ordering, inverted ordering and quasi-degeneracy.  We find that when the constraints of DM relic density and comic-ray antiproton spectrum are taken into account, the observed excess of positron/electron flux could be explained well in normal ordered neutrino mass spectrum. Moreover, excess of comic-ray neutrinos is implied in our model. We find that our results on $\langle \sigma v \rangle$ are satisfied with and close to the upper limit of IceCube analysis. More data from comic-ray neutrinos could test our model. 

\end{abstract}

\maketitle


\section{Introduction} 

Two strong direct evidences indicate the existence of new physics: one is the observations of neutrino oscillations, which lead to massive neutrinos~\cite{PDG2012}, and another one is the astronomical evidence of dark matter (DM), where a weakly interacting massive particle (WIMP) is the candidate in particle physics. 
The Planck best-fit  for  the DM  density,  which combines the data of
WMAP polarization at low multipoles, high-$\ell$ experiments and baryon acoustic oscillations (BAO), etc., 
 now is given by 
\cite{Ade:2013ktc}
 \be
 \Omega h^2=0.1187\pm0.0017\,. \label{eq:omega}
 \ed
Until now, we have not concluded what the DM is and what the masses of neutrinos originate. 
It is interesting if we can accommodate both  DM issue and neutrino masses in the same framework. 

Although the probe of  DM  could be through the direct detection experiments, however
%
according to the recent measurements by  LUX Collaboration~\cite{Akerib:2013tjd} and XENON100~\cite{XENON100}, we are still short of clear signals  and the cross section for elastic scattering of nuclei and DM has been  strictly limited.   
In contrast, the potential DM signals  have been observed by the indirect detections. For instance, the recent results measured by AMS-02~\cite{Aguilar:2013qda} have confirmed the excess of positron fraction which was observed earlier by PAMELA~\cite{Adriani:2008zr} and Fermi-LAT~\cite{FermiLAT:2011ab} experiments.
 %
Additionally, the excess of positron+electron flux above the calculated backgrounds is also observed by PAMELA~\cite{Adriani:2011xv}, Fermi-LAT~\cite{Ackermann:2010ij}, 
ATIC~\cite{Chang:2008aa} and HESS~\cite{Aharonian:2008aa, Aharonian:2009ah}.
Inspired by the observed anomalies, various interesting possible mechanisms to generate the high energy positrons and electrons are proposed, such as pulsars \cite{pulsar1,pulsar2}, dark matter annihilations \cite{DManni,Baek:2008nz,Ko:2010at} and dark matter decays \cite{Baek:2014goa,DMdecay,Chen:2009mf}.

The origin of neutrino masses is one of most mysterious problems in high energy physics. Before nonzero neutrino masses were found, numerous mechanisms had been proposed to understand the source of neutrino masses, such as type-I seesaw \cite{SeeSaw} and type-II seesaw \cite{ Magg:1980ut,Konetschny:1977bn} mechanisms, where the former introduced the heavy right-handed neutrinos and the latter extended the standard model (SM) by including a $SU(2)$ Higgs triplet.  
Since the triplet scalars only couple to leptons, based on this character, it may have interesting impacts on the cosmic-ray positrons, electrons and neutrinos. We therefore study a simple extension of conventional type-II seesaw model by including the possible DM effects.  

For studying the excess of cosmic rays by DM annihilation and the masses of neutrinos, 
we add an extra Higgs doublet $(\Phi)$ and a Higgs triplet ($\Delta$) to the SM.  Besides the gauge symmetry $SU(2)_L\times U(1)_Y$, in order to get a stable DM, we impose a discrete $Z_2$ symmetry in our model, where the $\Phi$ is  $Z_2$-odd and the $\Delta$ and SM particles are $Z_2$-even. The $Z_2$ odd doublet is similar to the one in inert Higgs doublet (IHD) model~\cite{Ma:2006km,Barbieri:2006dq},
where the IHD model has been studied widely in the literature, such as DM direct detection \cite{Barbieri:2006dq,LopezHonorez:2006gr,Arhrib:2013ela}, cosmic-ray gamma spectrum~\cite{Gustafsson:2007pc}, cosmic-ray positrons and antiproton fluxes~\cite{Nezri:2009jd}, collider signatures~\cite{Cao:2007rm,Dolle:2009ft}, etc.  
The lightest neutral odd particle could be either CP-odd or CP-even,  in this work we will adopt the CP-even boson as the DM candidate. 
 For explaining the observed excess of cosmic rays, 
 we set the odd particle masses  at TeV scale. 
 
There are two motivations to introduce the Higgs triplet. First, like the type-II seesaw mechanism \cite{Magg:1980ut,Konetschny:1977bn}, the small neutrino masses could be explained by the small VEV of triplet without introducing heavy right-handed neutrinos. Second,  the  excess of cosmic-ray  appears in positrons and electrons, however, by the measurements of  AMS~\cite{Aguilar:2002ad}, PAMELA~\cite{Adriani:2010rc} and HESS~\cite{Asaoka:2001fv}, no excess is found in cosmic-ray antiproton spectrum.
Since triplet Higgs bosons interact with leptons but do not couple to quarks, it is interesting to explore if 
the observed excess  of positron fraction and positron+electron flux could be explained by the leptonic decays of Higgs triplet in DM annihilation processes. The model with one odd singlet  and one $SU(2)_L$ triplet has been studied and one can refer to Ref.~\cite{Dev:2013hka}. Furthermore, the search of doubly charged Higgs now is an important topic at colliders. 
If doubly charged Higgs is 100\% leptonic decays,  
  the experimental lower bound on its mass has been limited in the range between 375 and 409 GeV~\cite{Chatrchyan:2012ya, ATLAS:2012hi}.
  The detailed analysis and the implications at collider physics could consult  the Refs.~\cite{triplet search, Perez:2008ha,Arhrib:2011uy,delAguila:2013mia}. 

The decays of triplet particles  to leptons depend on the Yukawa couplings. As known, the Yukawa couplings could be constrained by the measured neutrino mass-squared differences and the mixing angles of Pontecorvo-Maki-Nakagawa-Sakata (PMNS) matrix \cite{Pontecorvo:1957cp,Maki:1962mu}, where the current data  are given by \cite{PDG2012}
 \be
 \Delta m^2_{21}&=&(7.50\pm 0.20)\times 10^{-5} eV^2 \,, \non \\
 | \Delta m^2_{31} | &=&  (2.32 ^{+0.12}_{-0.08} ) \times 10^{-3} eV^2\,, \non \\
 \sin^2(2\theta_{12}) &=& 0.857 \pm 0.024\,, \sin^2(2\theta_{23}) > 0.95\,, \non \\
   \sin^2(2\theta_{13})&=&0.095 \pm 0.01.
 \ed
 Since the data can not tell the mass pattern from various neutrino mass spectra,  in our study, we classify the mass spectra to be normal ordering (NO), inverted ordering (IO) and quasi-degeneracy (QD) \cite{PDG2012} and investigate their influence on the production of cosmic rays. Because we do not have any information on the Dirac $(\delta)$ and Majorana $(\alpha_{31, 21})$ phases in PMNS matrix,  we adopt four benchmark points that are used by CMS Collaboration for the search of doubly charged Higgs \cite{Chatrchyan:2012ya}. The  first three benchmark points stand for the NO, IO and QD with $\delta=\alpha_{31}=\alpha_{21}=0$ while the fourth one denotes the QD with $\delta=\alpha_{31}=0$ and $\alpha_{21}=1.7$.   We note that the necessary boost factor (BF) for fitting the measured cosmic-ray electron/positron flux by DM annihilation is regarded as astrophysical effects \cite{Kuhlen:2012ft}. We take the BF as a parameter and use the data of antiproton spectrum to bound it. 

Furthermore, since the singly charged and neutral triplet particles couple to neutrinos,  an excess of cosmic-ray neutrinos is expected in the model. We find that 
 a Breit-Wigner enhancement could occur at the production of neutrinos;
 therefore, without BF, a large neutrino flux from DM annihilation  could be accomplished.
Accordingly, with the same values of free parameters that fit the excess of cosmic-ray positron/electron flux, our results on neutrino excess from galactic halo could be close to the upper bound measured by IceCube~\cite{Abbasi:2011eq, Aartsen:2013mla}. 

The paper is organized as follows.
We introduce the gauge interactions of IHD and triplet, 
Yukawa couplings of triplet, 
and scalar potential in Sec II. 
The set of free parameters and the branching fractions of triplet particle decays are introduced  in Sec III.
In Sec IV, we discuss the constraints from relic density of DM and cosmic-ray antiproton spectrum. With the values of constrained parameters, we study  the fluxes of cosmic-ray positrons, electrons and neutrinos. %
We give a summary  in Sec V.

\section{Inert Higgs doublet in type-II seesaw model}

In this section, we introduce the new interactions in the model. In order to have a stable DM, we consider the symmetry $SU(2)_L\times U(1)_Y \times Z_2$.  For generating the masses of neutrinos and having a DM candidate, we extend the SM to include a scalar triplet $\Delta$ with hypercharge $Y=2$ and a scalar doublet $\Phi$ with hypercharge $Y=1$. The SM particles and the triplet $\Delta$ are $Z_2$-parity even  while the new doublet $\Phi$ is $Z_2$-parity odd. 
Since the DM does not decay in the model, therefore, $\Phi$ cannot develop a VEV when electroweak symmetry is broken. 

The couplings in the SM are well known, therefore we do not further discuss them. With the new $Z_2$-parity,  the involved new gauge interactions, new Yukawa couplings and scalar potential are written as 
 \be
 {\cal L}_{NP} =( D_\mu \Phi)^{\dagger} D^\mu \Phi + (D_\mu \Delta )^\dagger D^\mu \Delta- \left[  \frac{1}{2}L^T C({\bf y + y^T})  i\sigma_2 \Delta P_L L +h.c. \right] - V(H, \Phi, \Delta)\,,
  \label{eq:lang_np}
 \ed
 where we have suppressed the flavor indices in Yukawa sector, ${\bf y}$ denotes the $3\times 3$ Yukawa matrix, $P_L=(1-\gamma_5)/2$, $L^T=(\nu_\ell, \ell)$ is the lepton doublet, $\sigma_2$ is the second Pauli matrix and $C=i\gamma_0 \gamma_2$. Due to $\Phi$ being $Z_2$ odd, it cannot couple to SM fermions. 
The representations for SM Higgs doublet H, $\Phi$ and  triplet are chosen as
 \be
H&=&\left( \begin{array}{c}
   G^+ \\
   (v_0+ h+i G^0)/\sqrt{2} \\
     \end{array} \right)\,, ~~~ \Phi=\left( \begin{array}{c}
   H^+ \\
   (S+iA)/\sqrt{2} \\
     \end{array} \right)\,, \non \\
     \Delta &=&  \left( \begin{array}{c} 
     \delta^{++}  \\ 
     \delta^+ \\
    (v_\Delta + \delta^0 + i\eta^0)/\sqrt{2}  \end{array} \right) ~ \text{or} ~\left( \begin{array}{cc} 
    \delta^+/\sqrt{2} & \delta^{++}  \\ 
    (v_\Delta + \delta^0 + i\eta^0)/\sqrt{2} & -\delta^+/\sqrt{2} \\ 
 \end{array}  \right)\,,
 \ed
where $v_{0(\Delta)}$  are the VEV of neutral component of $H(\Delta)$ and their values are related to the parameters of scalar potential.  There are two ways to present $\Delta$: for gauge interactions we use $3\times 1$ column vector but  for Yukawa couplings and scalar potential, we use $2\times 2$ matrix. Since the mixing of $H$ and $\Delta$ is related to the  small $v_\Delta$, which is constrained by $\rho$ parameter and the masses of neutrinos, the Goldstone bosons and Higgs boson are mainly from the SM Higgs doublet. Below we discuss each sector individually.  

\subsection{ Gauge Interactions  } 

The covariant derivatives for scalar doublet and triplet could be expressed by 
 \be
 D_\mu = \partial_\mu + i \frac{g}{\sqrt{2}} \left( { \bf T^+} W^+_\mu + {\bf T^-} W^{-}_\mu \right) + i\frac{g}{c_W} \left(  {\bf T^3} -  s^2_W {\bf Q} \right) Z_\mu+ i e {\bf Q} A_\mu \,.
 \ed
The $W^{\pm}_\mu$, $Z_\mu$ and $A_\mu$ stand for the gauge bosons in the SM, $g$ is the gauge coupling of $SU(2)_L$ and $s_W(c_W)=\sin\theta_W (\cos\theta_W)$ with $\theta_W$ being the Weinberg angle.  For scalar doublet, ${\bf T^{\pm}}=(\sigma^1 \pm i \sigma^2)/2$ and ${\bf T^{3}}=\sigma^3$ are associated with Pauli matrices and diag${\bf Q}=(1, 0)$ is the charge operator. For scalar triplet, the charge operator is diag${\bf Q}=(2, 1, 0)$, ${\bf T^{\pm}}=T_1 \pm i T_2$ and the generators of $SU(2)$ are set to be 
 \be
  T_1=\frac{1}{\sqrt{2}}\begin{pmatrix}
  0 & 1 & 0 \\
   1 & 0 & 1 \\
    0 & 1 & 0 \\
  \end{pmatrix}\,,
~ T_2= \frac{1}{\sqrt{2}} \begin{pmatrix}     0 & -i & 0 \\
   i & 0 & -i \\
    0 & i & 0 \\
     \end{pmatrix}\,,~  T_3= \begin{pmatrix}
   1  & 0 & 0 \\
   0  & 0 &  0 \\
   0  & 0 & -1 \\
     \end{pmatrix}\,. 
 \ed
The kinetic terms of SM Higgs and $\Delta$ will contribute to the masses of $W^\pm$ and $Z$ bosons. After spontaneous symmetry breaking (SSB), the masses of $W^\pm$ and $Z$ bosons are given by
 \be
 m^2_W &=& \frac{g^2 v^2_0}{4} \left( 1 + \frac{2v^2_\Delta }{v^2_0} \right) \,, \non \\
 m^2_Z &=& \frac{g^2 v^2_0}{4\cos^2\theta_W} \left( 1 + \frac{4 v^2_\Delta}{v^2_0 }\right)\,.
 \ed
As a result, the $\rho$-parameter at tree level could be obtained as
 \be
 \rho= \frac{m^2_W}{m^2_Z c^2_W} = \frac{ 1+ 2 v^2_\Delta/ v^2_0} { 1+4 v^2_\Delta/v^2_0}\,.
 \ed
Taking the current precision measurement for $\rho$-parameter to be $\rho=1.0004^{+0.0003}_{-0.0004}$ \cite{PDG2012}, we get  $v_{\Delta} < 3.4$ GeV when 2$\sigma$ errors is taken into account.  

We find that the gauge interactions of triplet particles such as $\delta^0 W^{+}W^-$ and  $\delta^0 ZZ$, which will be directly related to the relic density and excess of cosmic rays, are all proportional to $v_\Delta$. 
 For small $v_\Delta$, the effects are negligible.
It is known that triplet particles have two main decay channels: one is decaying to paired gauge bosons and the other is leptonic decays. In order to obtain the excess of cosmic-ray positrons/electrons and avoid getting a large cosmic-ray antiproton spectrum, the paired gauge boson channel should be suppressed. For achieving  the purpose, we take $v_\Delta < 10^{-4}$ GeV \cite{Perez:2008ha}. Although the vertex of $ZY\bar Y$ with $Y=\delta^{(++,+)}$ is important to produce the pair of triplet particles, however the (co)annihilation through the couplings of $H^+ H^- A$, $H^+ H^- Z$ and  $SAZ$ is suppressed by the low momenta of odd particles. Therefore,  their effects are not significant.

In order to satisfy the measured relic density $\Omega h^2$ and produce interesting excess of cosmic rays, the important gauge interactions are only associated with odd particles.  The relevant interactions are written as 
\be
{\cal L}_G&=&- \frac{g}{2}\left( S ( p_{S}- p_{H^-})^\mu + i A (p_A - p_{H^-})^\mu \right) W^{+}_{\mu} H^- 
 -i \frac{ g}{2\cos\theta_W}  ( p_S - p_A)^\mu Z_\mu A S  \non  \\
&&+\left[ \frac{g}{2} (S-iA) H^+ W^-_\mu \left(eA^\mu + \frac{g\sin^2\theta_W}{\cos^2\theta_W} Z^\mu\right) +h.c.\right] \non \\
&&+H^+ H^- \left( eA^\mu + \frac{g\cos2\theta_W}{\cos\theta_W} Z^\mu\right)^2 
+ \frac{g^2}{4} ( S^2 + A^2) W^{+}_\mu W^{-\mu}  \non \\
&& + \frac{g^2}{8\cos^2\theta_W} (S^2+A^2) Z_\mu Z^\mu\,. \label{eq:int_gauge}
\ed 
 According to Eq.~(\ref{eq:int_gauge}), we see that the DM (co)annihilation could produce $W^{+}W^{-}$ and $ZZ$ pairs by s- and t-channel. 
Although the $W^+ W^-$ and $ZZ$ pairs are open in the model and will contribute to the antiproton flux, however, we will see that the  produced antiprotons  in the  energy range of observations are still consistent with data measured by AMS~\cite{Aguilar:2002ad}, PAMELA~\cite{Adriani:2010rc} and HESS~\cite{Asaoka:2001fv}.

\subsection{ Yukawa Couplings} \label{sec:yukawa}
Next, we discuss the origin of neutrino masses and new lepton couplings in the model. Using the $2\times 2$ representation for $\Delta$, the Yukawa interactions in Eq.~(\ref{eq:lang_np}) could be decomposed as 
 \be
 -{\cal L}_{Y} &=& \frac{1}{2} \nu^T C {\bf h}P_L \nu \frac{v_\Delta + \delta^0 + i\eta^0 }{\sqrt{2}}
 -  \nu^T C {\bf h} P_L \ell  \frac{\delta^+}{\sqrt{2}}  \non \\
 &&- \frac{1}{2} \ell^T C {\bf h} P_L \ell \delta^{++} +h.c. 
 \label{eq:lang_Y}
  \ed
where ${\bf \bar h}=  {\bf y + y^T}$ and it is a symmetric $3\times 3$ matrix. 
 Clearly, the neutrino mass matrix is given by  ${\bf m_\nu }= v_\Delta {\bf h}/\sqrt{2}$. For explaining the tiny neutrino masses, we can adjust the  $v_\Delta$ and ${\bf h}$. In this paper, for suppressing the triple couplings of  triplet particle  and gauge bosons
 so that the leptonic triplet decays are dominant, we adopt  $v_{\Delta} < 10^{-4}$ GeV \cite{Perez:2008ha}.   By using PMNS matrix \cite{Pontecorvo:1957cp,Maki:1962mu}, the Yukawa couplings could be determined by the neutrino masses and the elements of PMNS matrix. The relation is given by
 \begin{equation}
\label{eq:yll}
{\bf h}= \frac{\sqrt{2}}{v_{\Delta}} U^*_{\rm PMNS} {\bf m}^{\rm dia}_{\nu}U^{\dagger}_{\rm PMNS} \,,
 \end{equation}
where ${\bf m}^{\rm dia}_\nu = {\rm diag}( m_1, m_2 , m_3 )$, $m_i$s are the physical masses of neutrinos and PMNS matrix is parametrized by \cite{PDG2012}
\begin{equation}
\label{MNS}
U_{\rm PMNS} = \begin{pmatrix} 
c_{12} c_{13} & s_{12} c_{13} & s_{13} e^{-i \delta} \\
-s_{12} c_{23} - c_{12} s_{23} s_{13} e^{i \delta} & c_{12} c_{23} -s_{12} s_{23} s_{13} e^{i \delta} & s_{23} c_{13} \\
s_{12} s_{23} - c_{12} c_{23} s_{13} e^{i \delta} & -c_{12} s_{23} - s_{12} c_{23} s_{13} e^{i \delta} & c_{23} c_{13} 
\end{pmatrix} \times \text{diag}(1, e^{i \alpha_{21}/2}, e^{i\alpha_{31}/2})
\end{equation}
with $s_{ij} \equiv \sin \theta_{ij}$, $c_{ij} \equiv \cos \theta_{ij}$ and $\theta_{ij}=[0, \pi/2]$. $\delta=[0,\pi]$ is the Dirac CP violating phase and $\alpha_{21, 31}$ are  Majorana CP violating phases.   According to Eq.~(\ref{eq:lang_Y}),  the couplings of triplet particles to SM leptons are all related to ${\bf h}$, therefore the Yukawa couplings of $\delta^{\pm\pm}$, $\delta^{\pm}$ and $\delta^0 (\eta^0)$  are limited by the neutrino experiments. If we set $v_{\Delta} h_{\ell' \ell}/\sqrt{2} = m_{\ell' \ell}$, the Eq.~(\ref{eq:yll}) could be decomposed as 
\begin{eqnarray}
m_{ee} &=& m_1 (c_{12} c_{13})^2 + m_2 e^{-i\alpha_{21}} (s_{12} c_{13})^2 + m_3 e^{-i(\alpha_{31}-2 \delta)} s_{13}^2  \non \\
m_{\mu \mu} &=& m_1 (s_{12} c_{23}+c_{12} s_{23} s_{13} e^{-i \delta})^2 +m_2 e^{-i \alpha_{21}} (c_{12}c_{23}-s_{12}s_{23}s_{13}e^{-i\delta})^2 + m_3 e^{-i \alpha_{31}} (s_{23}c_{13})^2  \non \\
m_{\tau \tau} &=& m_1 (s_{12}s_{23}-c_{12}c_{23}s_{13} e^{-i\delta})^2 + m_2 e^{-i \alpha_{21}} (c_{12}s_{23}+s_{12}c_{23}s_{13}e^{-i\delta})^2 + m_3 e^{-i \alpha_{31}} (c_{23} c_{13})^2 \non \\
m_{e \mu} &=& -m_1 c_{12} c_{13} (s_{12} c_{23}+c_{12} s_{23} s_{13} e^{-i \delta})+m_2 e^{-i \alpha_{21}} s_{12} c_{13} (c_{12} c_{23}-s_{12} s_{23} s_{13}e^{-i \delta}) \nonumber \\
&& + m_3 e^{-i(\alpha_{31}-\delta)} s_{23} s_{13} c_{13}  \non \\
m_{e \tau} &=& m_1 c_{12} c_{13} (s_{12}s_{23}-c_{12}c_{23}s_{13}e^{-i\delta})-m_2 e^{-i \alpha_{21}} s_{12} c_{13} (c_{12} s_{23}+s_{12}c_{23}s_{13}e^{-i\delta}) \nonumber \\
&& + m_3 e^{-i(\alpha_{31}-\delta)} c_{23} s_{13} c_{13}  \non \\
m_{\mu \tau} &=& -m_1 (s_{12}c_{23}+c_{12}s_{23}s_{13}e^{-i \delta})(s_{12}s_{23}-c_{12}c_{23}s_{13}e^{-i \delta}) \nonumber \\
&& -m_2 e^{-i\alpha_{21}} (c_{12}c_{23}-s_{12}s_{23}s_{13}e^{-i \delta})(c_{12}s_{23}+s_{12}c_{23}s_{13}e^{-i\delta}) \nonumber \\
&& + m_3 e^{-i \alpha_{31}} s_{23} c_{23} c_{13}^2\,. \label{eq:h}
\end{eqnarray}
  
As known that the neutrino experiments  can only measure the  mass squared difference between different neutrino species denoted by $\Delta m^2_{ij}=m^2_i - m^2_j$. 
 Since  the sign of $m^2_{31}$ and the absolute masses of neutrinos cannot be determined, this leads to three possible mass spectra in the literature and they are: \cite{PDG2012}\\
 \noindent (1) normal ordering (NO) ( $m_1<m_2<m_3$) with masses 
   \be
  m_{2(3)}= (m^2_1 + \Delta m^2_{21(31)})^{1/2}\,; \label{eq:no}
  \ed
\noindent (2) inverted ordering (IO) ( $m_3 < m_1 < m_2$) with masses 
    \be
  m_{1}= (m^2_3 + \Delta m^2_{13})^{1/2}\,, ~ m_2 = (m^2_1 + \Delta m^2_{21} )^{1/2}\,; \label{eq:io}
  \ed
  \noindent (3) quasi-degeneracy (QD)
   \be
   m_1 \simeq m_2 \simeq m_3 =m_0, ~ m_0 > 0.1 \text{~eV}.  \label{eq:qd}
   \ed 
Because the $m_{\ell' \ell}$ in Eq.~(\ref{eq:h}) depends on the masses of neutrinos,  the different mass patterns will lead to different patterns of Yukawa couplings.  Consequently, the branching fractions for ($\delta^{++}$, $\delta^{+}$, $\delta^0)$ decaying to leptons are also governed by the mass patterns.  We will explore their influence on the production of cosmic rays in DM annihilation.
 
\subsection{Scalar Potential}  
  
In the considered model,  one new odd doublet and one new triplet are included. Therefore, the new scalar potential with the $Z_2$-parity is given by
\be
V(H,\Phi,\Delta)&=& \mu^2 \Hd H + \lambda_1 ( \Hd H)^2 + m^2_\Phi \Pd \Phi + \lambda_2 ( \Pd \Phi)^2+\lambda_3 \Hd H \Pd \Phi + \lambda_4 \Hd \Phi \Pd H \non \\
& +& \frac{\lambda_5}{2}\left[ (\Hd \Phi)^2 +h.c. \right] + m^2_\Delta Tr\Dd \Delta + \mu_1 \left( H^T i\tau_2 \Dd H + h.c.\right) + \mu_2 \left( \Phi^T i\tau_2 \Dd \Phi \right) \non \\
&+& \lambda_6 \Hd H Tr\Dd \Delta + \bar\lambda_6 \Pd \Phi Tr\Dd \Delta 
+ \lambda_7 \Hd \Delta \Dd H 
+ \bar\lambda_7 \Pd \Delta \Dd \Phi + \lambda_8 \Hd \Dd \Delta H  \non \\
&+& \bar\lambda_8 \Pd \Dd \Delta \Phi + \lambda_9 (Tr\Dd \Delta)^2 + \lambda_{10} Tr(\Dd \Delta )^2\,. \label{eq:V}
\ed
Since the CP related issue is not discussed in this paper, all parameters are assumed to be real. Besides the SM parameters $\mu^2$ and $\lambda_1$,  there involve sixteen more free parameters. Since the SM Higgs doublet still dictates the electroweak symmetry breaking (EWSB),  $\mu^2 <0$ is required. The triplet and odd doublet have obtained their masses before EWSB, therefore we set $m^2_\Phi  (m^2_\Delta)>0$ and their values could be decided by the masses of odd (triplet) particles. We will see this point clearly later.

When $H$ and $\Delta$ develop the VEVs, the scalar potential as a function of $v_0$ and $v_\Delta$ is found as
\be
V(v_0,0, v_\Delta)= \frac{\mu^2}{2} v^2_0 + \frac{\lambda_1}{4} v^4_0 + \frac{m^2_\Delta}{2}v^2_\Delta+ \frac{\lambda_6 + \lambda_7}{4} v^2_0 v^2_\Delta + \frac{\lambda_9 + \lambda_{10}}{4} v^4_\Delta - \frac{\mu_1}{\sqrt{2}} v^2_0 v_\Delta\,. 
\ed
By using the minimal conditions of $\partial V/\partial v_0=0$ and $\partial V/\partial v_\Delta=0$, we easily get 
\be
v_0 \approx \sqrt{\frac{-\mu^2}{\lambda_1}}\,, v_{\Delta} \approx \frac{\mu_1 v^2_0}{\sqrt{2} \left(m^2_\Delta + \frac{\lambda_6 + \lambda_7}{2} v^2_0 \right)}\,, \label{eq:VEV}
\ed
where the condition of  $v_\Delta \ll v_0$ has been used. From the results, we see that $\mu_1$ is proportional to 
$v_{\Delta}$. In the scheme of $\mu_1\sim v_\Delta < 10^{-4}$ GeV,   the associated effects in DM annihilation could be ignored. 

According to Eq.~(\ref{eq:V}), we can obtain the masses of new scalars and triple (quadratic) interactions of odd particles and triplet particles. We first discuss the masses of new scalar particles. For odd particles,  if the small $v_{\Delta}$ effects are ignored, we find that the masses of $(S, A, H^\pm)$ are the same as those in the IHD model \cite{Ma:2006km,Barbieri:2006dq} and given by 
\be
m^2_S= m^2_\Phi + \lambda_L v^2_0\,, ~m^2_A - m^2_S = -\lambda_5 v^2_0\,, ~m^2_{H^\pm}=m^2_\Phi + \frac{\lambda_3} {2}v^2_0
\ed
with $\lambda_L=(\lambda_3+\lambda_4+\lambda_5)/2$. For SM Higgs and triplet particles, although the mixture of H and $\Delta$ could arise from triple coupling $\mu_1$ term and quadratic terms, however, they are all related to $v_{\Delta}$ and $\mu_1$. Due to $\mu_1\sim v_\Delta$, the mixing effects could be neglected.  For illustrating the small mixing effect, we take $G^0-\eta^0$ as the example. According to Eq.~(\ref{eq:V}), the mass matrix for $G^0-\eta^0$ is given by
\be
M_{G^0 \eta^0}=\begin{pmatrix}
                   \mu^2+ \lambda_1 v^2_0 + \sqrt{2} \mu_1 v_\Delta + \frac{\lambda_6}{4} v^2_\Delta & \sqrt{2} \mu_1 v_0 \\
                   \sqrt{2}\mu_1 v_0 & m^2_\Delta + \frac{\lambda_6 + \lambda_7}{2} v^2_0 + v^2_\Delta (\lambda_9 + \lambda_{10})
\end{pmatrix}\,.
\ed
If we take $\mu_1, v_\Delta \ll v_0$, the mixing angle of $G^0$ and $\eta^0$ is  $\theta_{G^0\eta^0}\sim 2v_\Delta/v_0\ll 1$. With Eq.~(\ref{eq:VEV}) and ignoring the small effects, we get
 \be
 m^2_{G^0} &\approx & \mu^2 + \lambda_1 v^2_0 \approx 0\,, \non \\
 m^2_{\eta^0} & \approx & m^2_\Delta + \frac{\lambda_6 + \lambda_7}{2} v^2_0\,.
 \ed
Similarly, other scalar mixings are also small and negligible. Consequently, 
 the masses of SM Higgs and triplet particles are given by
 \be
 m^2_h &\approx& 2\lambda_1 v^2_0\,, m^2_{\delta^0} \approx m^2_{\eta^0}\approx m^2_{\Delta} + \frac{\lambda_6 + \lambda_7}{2} v^2_0\,, \non \\
m^2_{\delta^{++}}&\approx& m^2_\Delta + \frac{\lambda_6 + \lambda_8}{2} v^2_0\,,\non\\
m^2_{\delta^{+}}&\approx&  \frac{1}{2}( m^2_{\delta^{++}}+ m^2_{\delta^0})\,.
\ed
We see that $m_{\delta^0}\approx m_{\eta^0}$ and $m_{\delta^+}$ is fixed when $m_{\delta^0}$ and $m_{\delta^{++}}$ are determined. We point out that the vertices from  $\Phi^T i\sigma_2 \Delta^\dagger \Phi$ term are associated with a dimensional parameter $\mu_2$. Unlike $\mu_1$ which is limited to be much smaller than $v_0$, the value of $\mu_2$ could be as large as few hundred GeV. It will have an interesting contributions to the cosmic-ray flux of neutrino from DM annihilation. We will further discuss its effects later. 

Now we discuss the triple and quadratic interactions of scalar particles that are responsible for relic density and production of cosmic rays.  According to Eq.~(\ref{eq:V}),  there appear lots of new interactions among new scalar particles. However, for explaining the measured relic density and studying the excess of cosmic rays by the DM (co)annihilation, here we display those relevant interactions in Table~\ref{tab:34}.  In the table, we have ignored the couplings related to $v_\Delta$ and $\mu_1$. 
\begin{table}[tbhp]
\caption{Triple and quadratic couplings of $Z_2$-odd and -even scalar  particles for relic density and the excess of cosmic rays.  } 
\begin{ruledtabular}
\begin{tabular}{cc|cc} 
 Vertex & Coupling &  Vertex & Coupling \\ \hline
  $SSh$ & $2\lambda_L v_0$  & AAh & $(\lambda_3 + \lambda_4 -\lambda_5)  v_0$ \\ \hline
  $SS (AA) \delta^0$ & $\mp \sqrt{2} \mu_2$ & $SA\eta^0$ & $-\sqrt{2} \mu_2$ \\ \hline
  $SH^{\mp}\delta^{\pm}$ & $-\mu_2$ & $A H^{\mp} \delta^{\pm}$ & $\pm i \mu_2$ \\ \hline
  $H^+ H^- h$ & $\lambda_3 v_0$ & $H^\pm H^\pm \delta^{\mp\mp}$ & $2\mu_2$ \\ \hline
   $\delta^+ \delta^{-} h$ & $(\lambda_6 + (\lambda_7 + \lambda_8)/2) v_0$ &  $\delta^{++} \delta^{--} h$ & $(\lambda_6 + \lambda_8)v_0$ \\ \hline 
     $\delta^{0} \delta^0 (\eta^0 \eta^0) h$ & $(\lambda_6 + \lambda_7) v_0$    & $hhh$ & $6\lambda_1 v_0$ \\ \hline \hline
  $SS(H^+ H^-)hh$ & $2\lambda_L (\lambda_3)$ & $AAhh$ & $\lambda_3+\lambda_4-\lambda_5$ \\ \hline
  $SS(AA)\delta^+ \delta^-$ & $\bar \lambda_6 + (\bar\lambda_7 +\bar\lambda_8)/2$ & $(S^2,A^2) \delta^{0}\delta^0 [\eta^0 \eta^0]$ & $\bar\lambda_6 + \bar\lambda_7$ \\ \hline
$SS(AA)\delta^{++}\delta^{--}$ & $\bar\lambda_6 + \bar\lambda_8$  & $H^+ H^- \delta^+ \delta^-$ & $ \bar\lambda_6 + (\bar\lambda_7 + \bar\lambda_8)/2$  \\ \hline
$H^+ H^- \delta^{++} \delta^{--}$ & $\bar\lambda_6 + \bar\lambda_7$ & $H^+ H^- \delta^0 \delta^0 ( \eta^0 \eta^0)$ & $\bar\lambda_6 + \bar\lambda_8 $ \\ \hline
 $SH^- \delta^- \delta^{++} ( H^+ \delta^+ \delta^{--} )$ & $ -(\bar\lambda_7 - \bar\lambda_8)/2$ & $A H^+ \delta^{+} \delta^{--}(H^{-} \delta^- \delta^{++} )$ & $\pm i/2 (\bar\lambda_7 - \bar\lambda_8)$  \\ \hline
 $S H^{\mp} \delta^{\pm}  \delta^{0} $ & $(\bar\lambda_7 - \bar\lambda_8)/(2\sqrt{2})$ &  $AH^{\mp} \delta^{\pm}  \eta^0 $ & $(\bar\lambda_7 - \bar\lambda_8)/(2\sqrt{2})$ \\ \hline
$SH^{\pm} \delta^{\mp} \eta^0$ & $ \pm i/(2 \sqrt{2}) (\bar\lambda_7 - \bar\lambda_8)$ & $AH^\mp \delta^\pm \delta^0$ & $\pm i/(2\sqrt{2}) (\bar\lambda_7 - \bar\lambda_8)$
  \end{tabular}
\end{ruledtabular}
\label{tab:34}
\end{table}%

\section{Setting parameters and Branching fractions of triplet decays }

In the model, there involve many new free parameters and some of them are not independent. Since we are interested in the mass dependence, we will take the masses as the variables. Hence,  the set of new independent free parameters can be chosen as follows:
 \be
 \{ m^2_S, m^2_A, m^2_{H^\pm},  \lambda_2, \lambda_L, m^2_{\delta^0}, m^2_{\delta^{\pm\pm}}, \mu_2, \lambda_6, \xi_A, \bar\lambda_6, 
  \chi_A,  \chi_B, \lambda_9, \lambda_{10} \} \label{eq:para}
 \ed
with $\xi_{A(B)}=\lambda_6 + \lambda_{8(7)}$ and $\chi_{A(B)}=\bar\lambda_6 + \bar\lambda_{8(7)}$. Accordingly, the decided parameters are expressed as
 \be
 m^2_\Phi&=&m^2_S - \lambda_L v^2_0\,,~ \lambda_3= \frac{2}{v^2_0} \left( m^2_{H^\pm} -m^2_\Phi \right)\,,\non \\
 \lambda_5&=& \frac{m^2_S - m^2_A }{v^2_0}\,, ~\lambda_4 = 2\lambda_L - \lambda_3 - \lambda_5\,, \non \\
 m^2_\Delta &=& m^2_{\delta^{\pm\pm}} - \frac{\xi_A}{2} v^2_0\,, ~\xi_B=\frac{2}{v^2_0}\left( m^2_{\delta^0} - m^2_\Delta \right)\,, \non \\
 \lambda_7 &=& \xi_B-\lambda_6\,,~\lambda_8 = \xi_A - \lambda_6\,, \non \\
  \bar\lambda_7 &=& \chi_A - \bar\lambda_6\,,~\bar\lambda_8=\chi_B-\bar\lambda_6\,.
 \ed
In our approach, we choose the lightest odd particle (LOP) to be $S$, the DM candidate. 
For revealing the triplet contributions to the production of cosmic rays and relic density of DM,
we suppress the effects from the original IHD model by assuming $\lambda_L=0$ and the mass differences of odd particles being within few GeV, where the former  leads the vertex of  Higgs-$SS(AA)$ to vanish and the latter inhibits the (co)annihilation processes induced by gauge interactions.
%
Therefore, for simplifying our numerical analysis, the values of free parameters are adopted as follows:
 \be
 \lambda_L=0\,,~m_S &=& m_A - 1~\text{GeV}\,, m_{H^\pm}=m_A\,, \non \\
 m_{\delta^{\pm\pm}} &=& m_{\delta^{\pm}}=m_{\delta^{0}} \equiv m_\delta =500~\text{GeV}\,.
 \label{eq:parameter_setting}
 \ed
 Due to  the mass degeneracy in triplet particles, i.e. $\lambda_{6}+\lambda_{7(8)}$ =0, the interactions of $(\delta^{++}\delta^{--}, \delta^{+} \delta^{-}, \delta^{0} \delta^{0}, \eta^{0} \eta^{0})h$ in Table~\ref{tab:34} also vanish. 
%

 The new source to produce the  positrons and neutrinos in the model is by the decays of  triplet particles, where
the main effects  are associated with Yukawa couplings. As mentioned in Sec. \ref{sec:yukawa}, although the Yukawa couplings have been constrained by neutrino experiments, however the neutrino mass spectrum,  Dirac phase and Majorana phases are still uncertain. For numerical analysis, we study three possible mass spectra defined in Eqs.~(\ref{eq:no})-(\ref{eq:qd}) by taking   $\delta=0$ and $\phi_{21(31)}=0$.
For comparison, we  also take $\delta=\phi_{31}=0$ and $\phi_{21}=1.7$ for QD to illustrate the influence of Majorana phase. Therefore, we use QDI and QDII to show the differences.

 Numerically,  we use the measured central values for  the mixing angles $(\theta_{ij})$ of PMNS matrix and for the mass squared differences, i.e.  the  inputs are taken as $\sin^2(2\theta_{12})= 0.857 $, $\sin^2(2\theta_{23}) =0.95$, $\sin^2(2\theta_{13})=0.095$, $m^2_{12}=7.5\times 10^{-5}$ eV and $|m^2_{31}|=2.32\times 10^{-3}$ eV.  For QD case, we set $m_0=0.2$ eV. As a result, the numerical values of $m_{\ell' \ell}$ in Eq.~(\ref{eq:h}) are given in Table~\ref{tab:v_mll}. Since we have taken $v_\Delta < 10^{-4}$ GeV, triplet particles mainly decay to leptons. Hence, the corresponding branching ratios (BRs) for triplet decays  are shown in Table~\ref{tab:brs}.  
 The values in brackets in Table~\ref{tab:brs} are the BRs for $\delta^{\pm}$ decays. The difference between $\delta^{\pm\pm}(\delta^0,\eta^0)$ and $\delta^{\pm}$ is because the former has two identical particles in the final state; therefore,  a proper symmetry factor has to be included. 
\begin{table}[htdp]
\caption{Values of Yukawa couplings (in units of $10^{-2}$ eV) in NO, IO, QDI and QDII. The abbreviations  could refer to the text.     }
\begin{ruledtabular}
\begin{tabular}{c|cccccc} 
 & $m_{ee}$ & $m_{e\mu}$ & $m_{e\tau}$ & $m_{\mu \mu}$ & $m_{\mu \tau}$ & $m_{\tau\tau}$ \\ \hline
NO & 0.3  & 0.6 & 0.1 & 1.4 & 1.0 & 1.4  \\ \hline
IO  &  1.5 & 0.7 & -1.1 & 1.0 & -1.3 & 1.6 \\ \hline
 QDI & 14.2 & 0 & 0 & 12.1 & -2.1 & 12.4 \\ \hline
 QDII& $10.3 e^{-i0.4}$ &  $5.9 e^{-i2.4}$ & $7.9e^{i0.7}$ & $8.8 e^{-i0.4} $ & $5.8 e^{i0.1} $ & $8.2 e^{-i0.9}$   
\end{tabular}
\end{ruledtabular}
\label{tab:v_mll}
\end{table}%
\begin{table}[bhtdp]
\caption{Branching ratios (BRs) for $\delta^{\pm\pm} (\delta^0, \eta^0)$ and  $\delta^{\pm}$  decays in NO, IO, QDI and QDII where the corresponding values of Yukawa couplings are given in Table~\ref{tab:v_mll}. $\ell_{f}$ could be charged lepton or neutrino and it depends on what its parent is.  The values in brackets denote the BRs for $\delta^{\pm}$ decays. }
\begin{ruledtabular}
\begin{tabular}{c|cccccc} 
 & $\ell_{e} \ell_{e}$ & $ \ell_{e} \ell_{\mu}$ & $\ell_{e} \ell_{\tau}$ & $\ell_{\mu} \ell_{\mu}$ & $\ell_{\mu} \ell_{\tau}$ & $\ell_{\tau} \ell_{\tau}$ \\ \hline
NO & 0.01 [0.02] & 0.1 [0.06]& 0[0] & 0.3 [0.39] & 0.29 [0.18] & 0.28 [0.35]  \\ \hline
IO  &  0.17 [0.23]& 0.08 [0.06]& 0.2 [0.14] & 0.08 [0.11] & 0.26 [0.17]& 0.21[0.29] \\ \hline
 QDI & 0.39 [0.40]& 0 [0]& 0 [0]& 0.29 [0.29]& 0.02  [0]& 0.30 [0.31] \\ \hline
 QDII& $0.21 [0.28]$ &  $0.13 [0.09]$ & $0.24 [0.16]$ & $0.15 [0.20]$ & $0.13 [0.09]$ & $0.13 [0.18]$   
\end{tabular}
\end{ruledtabular}
\label{tab:brs}
\end{table}%


\section{Relic density and fluxes and energy spectra of cosmic rays}

\begin{figure}[hpbt] 
\begin{center}
\subfigure[]{\includegraphics[width=50mm]{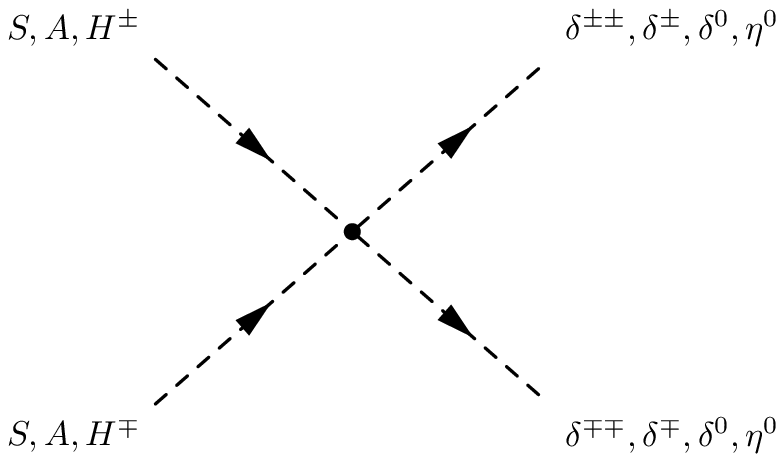} }
\subfigure[]{\includegraphics[width=50mm]{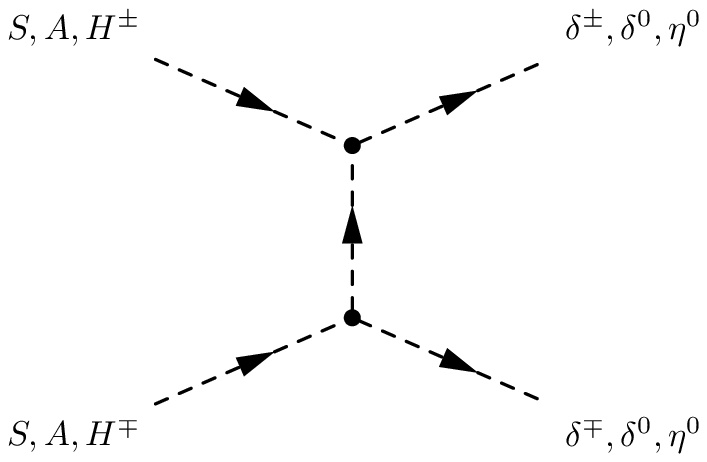}}
\subfigure[]{\includegraphics[width=50mm]{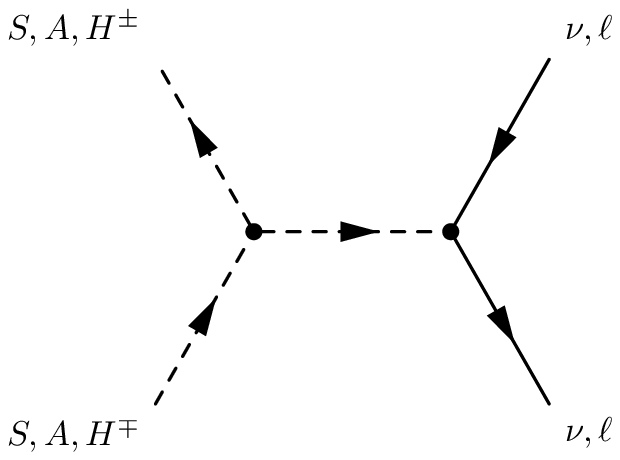}}
\caption{ Three scenarios of (co)annihilation processes of odd particles for relic density: (I) s-channel by quadratic interactions, (II) t-channel by triple interactions and (III) s-channel by triple interactions.  }
\label{fig:diagrams}
\end{center}
\end{figure}

After establishing our model and deciding the set of parameters, in this section we discuss 
 the effects of new interactions
on the relic density of DM and their implications on the indirect DM detection experiments for cosmic rays.
Since all odd particles belong to the IHD, the interesting 
(co)annihilation processes which determine the relic density of DM
could be classified as three scenarios: (I) s-channel from quadratic couplings, (II) t-channel from triple couplings and (III) s-channel from triple couplings, where the initial states  only  involve IHD  and the final states are triplet particles and leptons. The corresponding Feynman diagrams are shown in Fig.~\ref{fig:diagrams}. 
The particles  in initial and  final states are decided by the couplings  shown in  Table~\ref{tab:34}. 
Although $W^{\pm}$ and $Z$ pairs could be generated 
 by gauge interactions or by s-channel process mediated by SM-Higgs, due to the parameter setting in Eq.~(\ref{eq:parameter_setting}) and $m_\Phi \sim O({\rm TeV})$, their production cross sections by the (co)annihilation processes are secondary effects. We will ignore their contributions to relic density in our analysis.
%

In terms of the set of parameters in Eq.~(\ref{eq:para}) and the Yukawa couplings in Eq.~(\ref{eq:lang_Y}), the scenarios I, II and III depend on the parameter sets $\{\chi_A, \chi_B \}$, $\{ \mu_2 \}$ and $\{ \mu_2, h_{\ell' \ell} \}$, respectively. 
%
In addition, the two parameters $\chi_A$ and $\chi_B$ in scenario-I will result different energy spectra of positrons and neutrinos. 
 For further studying their contributions, we therefore consider three  schemes for scenario-I as follows : (I$_a$) $\chi_A \gg \chi_B\approx 0$,  (I$_b$)  $\chi_A = \chi_B$ and (I$_c$) $\chi_A \approx 0 \ll\chi_B$.

According to the taken values of parameters in Eq.~(\ref{eq:parameter_setting}) and Table~\ref{tab:v_mll}, the free parameters now are $m_S$, $\chi_{A,B}$, $\mu_2$ and $h_{\ell' \ell}$. 
We note that because of $h_{\ell' \ell}=\sqrt{2}m_{\ell' \ell}/ v_\Delta$, when the values of $m_{\ell'\ell}$ are fixed as shown in Table~\ref{tab:v_mll}, the associated free parameter of $h_{\ell' \ell}$ indeed is $v_\Delta$. 
 In our approach, we first constrain the free parameters so that the observed relic density of DM, $\Omega h^2$, could be explained in our chosen scenarios.
With the constrained parameters, we then estimate the 
fluxes of cosmic-ray antiprotons, positrons, electrons and neutrinos. As mentioned earlier, for explaining the excess of the cosmic rays by DM annihilation,  usually we need a BF. The BF could be arisen from several mechanisms, such as astrophysical origin \cite{Kuhlen:2012ft}, Sommerfeld enhancement~\cite{Sommerfeld}, near-threshold resonance and dark-onium formation~\cite{Dark-onium}, etc. Since we do not focus on such effect in the model, as used in the literature we take it as an undetermined parameter and its value could be limited by the antiproton flux measured by AMS~\cite{Aguilar:2002ad}, PAMELA \cite{Adriani:2010rc} and HESS~\cite{Asaoka:2001fv}. With the decided BF, we study the positron/electron and neutrino fluxes in various situations of free parameters.

\subsection{Relic Density}

Now we start to make the numerical analysis for the $\Omega h^2$ by DM $S$ (co)annihilation processes. 
For numerical calculations, we implement our model to CalcHEP \cite{CalcHEP} and use {\tt micrOMEGAs}~\cite{micrOMEGAs} to  estimate the $\Omega h^2$. To constrain the parameters, we require the relic density of $S$ to satisfy the 90\% CL (confidence level) range of its experimental value, written as
 \be
 0.1159\le \Omega h^2\le0.1215 \,.\label{eq:oh2}
  \ed
  In the following we individually discuss the contributions from  scenario-I, -II and -III.

\begin{figure}[hptb] 
\begin{minipage}{0.45 \hsize}
\begin{center}
\includegraphics[width=70mm]{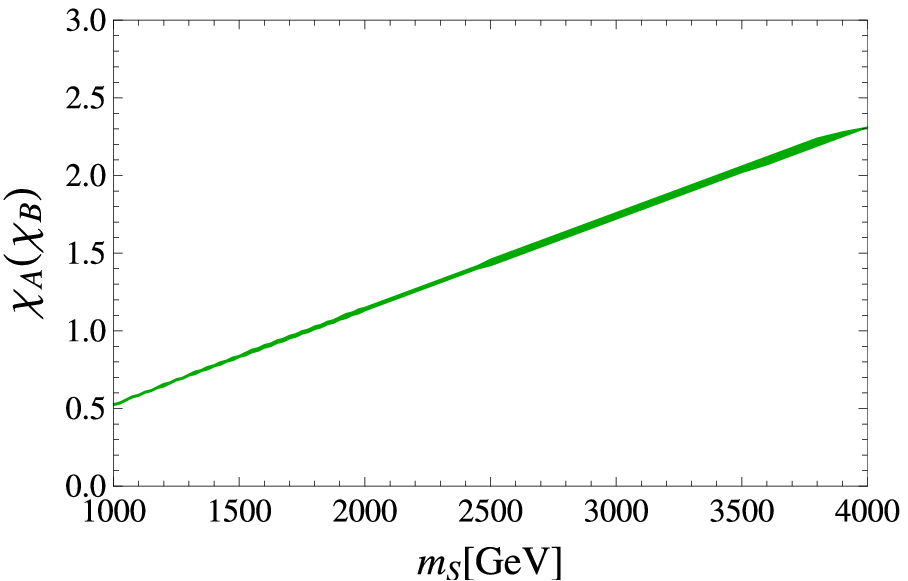} 
\end{center}
\end{minipage}
\begin{minipage}{0.45 \hsize}
\begin{center}
\includegraphics[width=70mm]{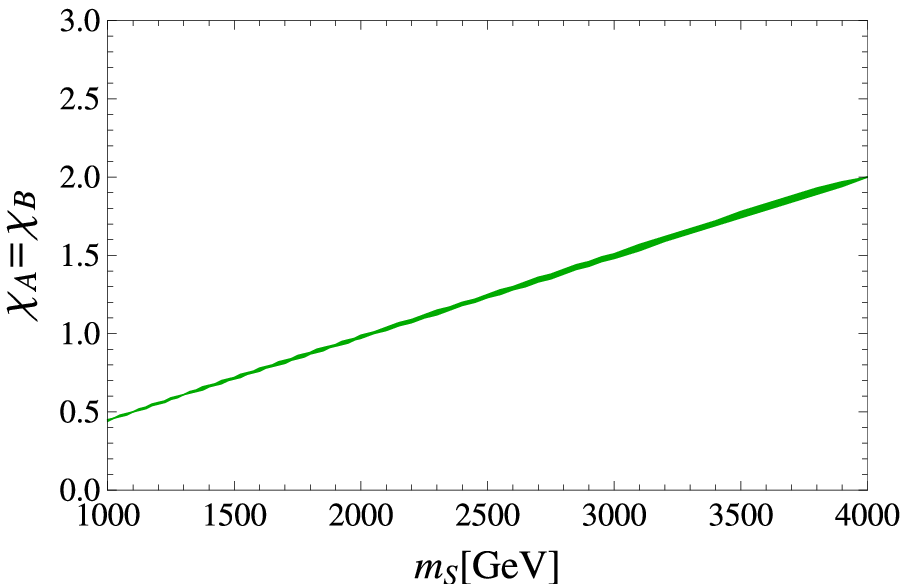}
\end{center}
\end{minipage} 
\caption{ The allowed region  in scenario-I when $\Omega h^2$ is satisfied.
Left panel denotes the correlation between $m_S$ and  $\chi_A(\chi_B)$ in scheme I$_{a(c)}$.  Right panel stands for the results of scheme I$_b$. Schemes I$_{a}$ and I$_{c}$ have the same results. 
\label{fig:allowed1}}
\end{figure}

For scenario-I, as stated above, 
we classify three schemes based on the relative magnitude of parameters $\chi_A$ and $\chi_B$.
With the values of parameters in Eq.~(\ref{eq:parameter_setting}) and the range of $\Omega h^2$ in Eq.~(\ref{eq:oh2}),  the correlation between $m_S$ and $\chi_{A}(\chi_{B})$  in scheme-I$_{a(c)}$ is given in the left panel of Fig.~\ref{fig:allowed1}.  We find that   the schemes I$_a$ and I$_c$ have the same contributions. 
Similarly, the results of  scheme I$_b$ are displayed in the right panel of Fig.~\ref{fig:allowed1}.  We see that $\chi_{A,B}$ of $O(1)$ can accommodate  to the observed $\Omega h^2$.
%

\begin{figure}[hpbt] 
\begin{minipage}{0.45 \hsize}
\begin{center}
\includegraphics[width=70mm]{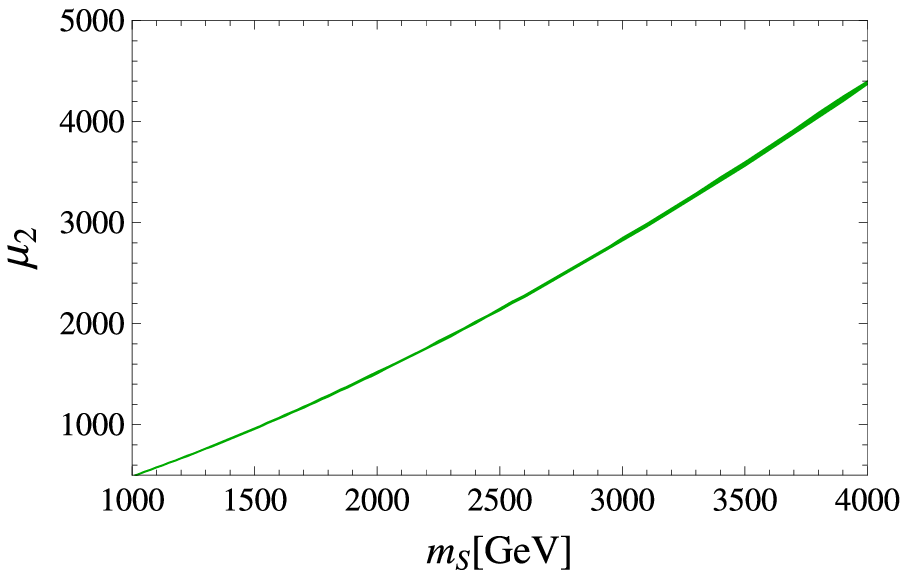} 
\end{center}
\end{minipage}
\begin{minipage}{0.45 \hsize}
\begin{center}
\includegraphics[width=70mm]{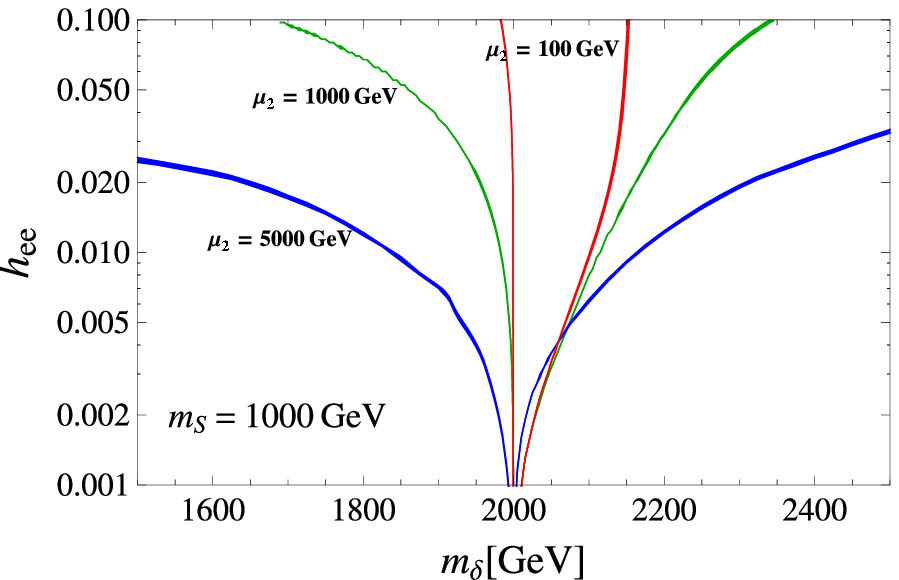}
\end{center}
\end{minipage} 
\caption{ The allowed region for scenario-II (left) and scenario-III (right). For scenario III, we plot the contours of $\Omega h^2$ as a function of $h_{ee}$ and $m_{\Delta}$ with various values of $\mu_2$ and $m_S=1000$ GeV. 
\label{fig:allowed2}}
\end{figure}

For scenario-II,  the  involved parameters are $m_S$  and $\mu_2$.
Differing from other parameters, $\mu_2$ is a mass dimension one parameter and its natural value could be from  GeV to TeV. With the data of $\Omega h^2$, we present 
the correlation between $m_S$ and  $\mu_2$   in the left panel of Fig.~\ref{fig:allowed2}.
By the plot, we see that 
if the scenario-II is the only source of the observed $\Omega h^2$,
the value of $\mu_2$ has to be as large  as the scale of $m_S$. 

For scenario-III, the related parameters are  $\mu_2$ and $h_{\ell' \ell} $. Thus, we expect that the limit on $\mu_2$ with $m_\Delta < m_S$ should be similar to the cases 
in scenario-II.
However, the more interest of this scenario is the reverse case. From Fig.~\ref{fig:diagrams}-III, we see that the intermediate state is triplet particle. When $m_\Delta \approx 2 m_S$ is satisfied, we have a large effect from the Breit-Wigner enhancement.  As discussed before, Yukawa couplings are determined by $h_{\ell' \ell}=\sqrt{2} m_{\ell' \ell}/v_\Delta$. If we use the fixed values in Table~\ref{tab:v_mll}, the free parameter is only $v_\Delta$. Since we only need one parameter to describe all $h_{\ell' \ell}$, here we just take $h_{ee}$ as the representative.   Once $h_{ee}$ is determined, other $h_{\ell' \ell}$ are also fixed. Accordingly, we present the results as a function of $h_{ee}$ and $m_\Delta$ with several values of $\mu_2$ in the right panel of Fig.~\ref{fig:allowed2}, where we only use the IO for neutrino mass spectrum and $m_S=1000$ GeV as an illustration. 
We find that due to the  Breit-Wigner enhancement, a smaller value of $\mu_2$  could get desired magnitude of the (co)annihilation cross section.

\subsection{ Antiproton spectrum and boost factor constraint}

As mentioned earlier, the necessary BF for 
explaining the excess of cosmic ray fluxes by DM annihilation is regarded as a parameter. 
 However, the value of BF can not be arbitrary  and we need to investigate the limit of BF by the observed data.
It is known that cosmic-ray antiprotons have been measured by AMS~\cite{Aguilar:2002ad}, PAMELA~\cite{Adriani:2010rc} and BESS~\cite{Asaoka:2001fv}. By the data, we see that below the energy of 100 GeV, the measurements fit well with the models of cosmic-ray background. Therefore, when the values of parameters are fixed by $\Omega h^2$, antiproton flux $\Phi_{\bar p}$ could provide an upper limit on the BF.

In the model, since triplet particles only couple to leptons and cannot produce the antiprotons, the channels to generate antiprotons are from $W$ and $Z$ decays following $WW $ and $ZZ$ pair production, where $WW$ and $ZZ$ are produced from s-channel and t-channel by the gauge interactions $SS(AA) VV$ and $S(A)H^\pm W^\mp [ SAZ ]$ that are listed in Table~\ref{tab:34}, respectively.  We note that  in our taken values of parameters, although the production of $WW$ and $ZZ$ for  contributing to  $\Omega h^2$ is secondary effects, however it becomes the leading contributions to the antiproton production. Additionally, since the gauge coupling is known and fixed, $m_S$ is the only free parameter for $WW$ and $ZZ$ production. Therefore, the limit on BF is clear. 

\begin{figure}[hpbt] 
\begin{minipage}{0.45 \hsize}
\begin{center}
\includegraphics[width=70mm]{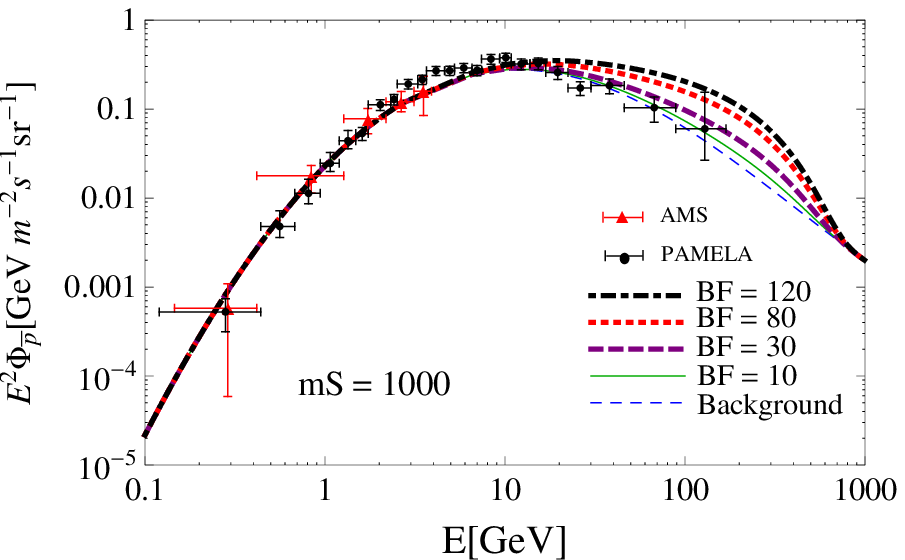} 
\end{center}
\end{minipage}
\begin{minipage}{0.45 \hsize}
\begin{center}
\includegraphics[width=70mm]{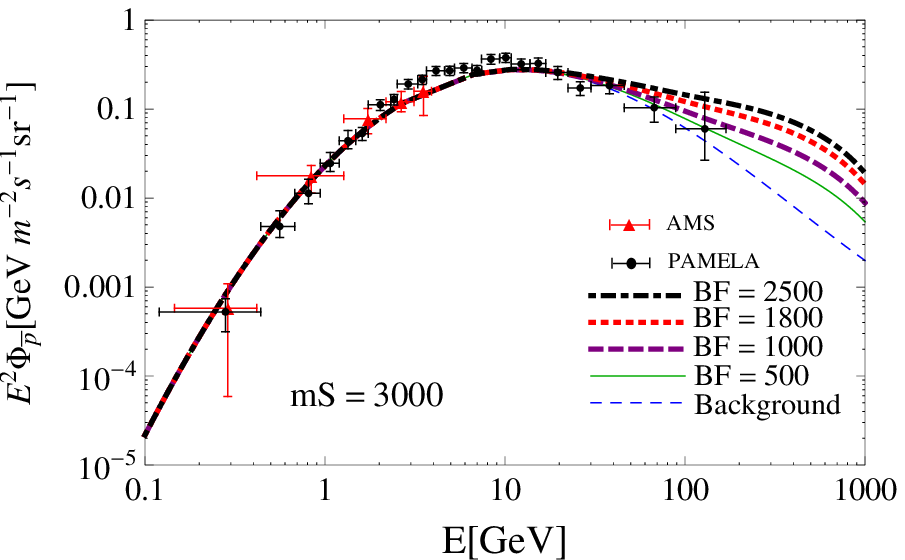}
\end{center}
\end{minipage} 
\caption{   Background and background + DM of cosmic-ray antiproton flux in different values of BF with $m_S=1000$ GeV (left) and $m_S=3000$ GeV (right). The data points stand for the PAMELA~\cite{Adriani:2010rc} and 
AMS~\cite{Aguilar:2002ad} results.
\label{fig:anti-proton}}
\end{figure}

For estimating $\Phi_{\bar p}$, we applied Navarro-Frenk-White (NFW) density profile for DM density distribution in the galactic halo~\cite{Navarro:1995iw}. For cosmic-ray antiproton background, we use the fitting function parametrized by \cite{Cirelli:2008id}
 \be
 log_{10} \Phi^{\rm bkg}_{\bar p} = -1.64 +0.07 x -x^2 -0.02 x^3 + 0.028 x^4
 \ed
 with $x= \log_{10} $T/GeV, where the result is arisen from the analysis in Ref.~\cite{Bringmann:2006im}. Using {\tt micrOMEGAs}, we present our results in Fig.~\ref{fig:anti-proton},  where the galactic propagation of charged particles and  solar modulation effect are also  taken into account. 
The left (right) panel of Fig.~\ref{fig:anti-proton} shows  the background and background$+$DM for $\Phi_{\bar p}$ in which $m_S=1000(3000)$ GeV is used and various values of BF are taken.  
Since PAMELA~\cite{Adriani:2010rc} and AMS~\cite{Aguilar:2002ad} results are more precise, we just show these data in the figure. We also use its data to constrain the BF. 
%
We find that the upper bound on the boost factor is  $\sim30(1800)$ for $m_S=1000(3000)$ GeV.
The upper value of BF with the corresponding value of  $m_S$ is given in Table~\ref{BoostFactor}.
By the results, we see clearly that the new physics contributions have a significant deviation from the background at $E> 100$ GeV. Therefore, the data at such energy region could test the model. 
\begin{table}[hpbt] 
\caption{Upper bound of the BF for different values of $m_S$.}
\begin{ruledtabular}
\begin{tabular}{ccccc} 
$m_S$ [GeV] & 1000 & 2000 & 3000 & 4000 \\ \hline
BF  & $ \lesssim 30$ & $ \lesssim 500$ & $\lesssim 1800$ & $ \lesssim 4500$   \\
\end{tabular} 
\end{ruledtabular}
 \label{BoostFactor}
\end{table}

\subsection{Cosmic-ray positron and electron spectra}

After discussing the constraints of free parameters and the limit of the BF, we investigate the influence of DM annihilation on the positron/electron and neutrino fluxes. We first study the case for cosmic-ray positrons/electrons. Besides the new source for the fluxes of electron and positron, we also need to understand the background contributions of primary and secondary electrons and secondary positrons, in which the former comes from supernova remnants and the spallation of cosmic rays in the interstellar medium, respectively, while the latter could be generated by primary protons colliding with other nuclei in the interstellar medium.
In our numerical calculations, we use the parametrizations, given by~ \cite{Baltz:1998xv, Baltz:2001ir}
\begin{align}
\Phi_{e^-}^{\rm prim}(E) &=  \kappa \frac{0.16 E^{-1.1}}{1+11 E^{0.9} + 3.2 E^{2.15}}  \quad  [ {\rm GeV^{-1} cm^{-2} s^{-1} sr^{-1}}], \non \\
\Phi_{e^-}^{\rm sec}(E) &=  \frac{0.70 E^{0.7} }{1+ 110 E^{1.5} + 600 E^{2.9} +580 E^{4.2}} \quad  [ {\rm GeV^{-1} cm^{-2} s^{-1} sr^{-1}}], \non \\
\Phi_{e^+}^{\rm sec}(E) &=  \frac{4.5 E^{0.7}}{1+650 E^{2.3}+1500 E^{4.2}} \quad  [ {\rm GeV^{-1} cm^{-2} s^{-1} sr^{-1}}], \label{eq:bkgrd}
\end{align} 
where $\Phi^{\rm prim(sec)}$ denotes the primary (secondary) cosmic ray. Accordingly, the total electron and positron fluxes are defined by
 \be
 \Phi_{e^-}&=& \kappa \Phi^{\rm prim}_{e^-} + \Phi^{\rm sec}_{e^-} + \Phi^{\rm DM}_{e^-}\,, \non \\
 \Phi_{e^+} &=& \Phi^{\rm sec}_{e^+} + \Phi^{\rm DM}_{e^+}\,,
 \ed
 where $\Phi^{\rm DM}_{e^{-(+)}}$ is the electron(positron) flux from DM annihilations.
According to Refs. \cite{Baltz:1998xv} and \cite{Moskalenko:1997gh}, we have regarded the normalization of the primary electron flux to be undetermined and parametrized by the parameter of $\kappa$.  In our analysis, we use $\kappa = 0.78$ to fit the experimental data for background.

When we study the relic density, we have used three scenarios to classify the parameters.  Except the scenario-III that only can contribute to cosmic-ray neutrinos, the scenario-I and -II could be also applied to the cosmic-ray electrons and positrons by  DM annihilation.   Since  the quadratic couplings of $SS\Delta \bar \Delta$ only depend on $\chi_A$ and $\chi_B$,  we could apply the schemes I$_a$, I$_b$ and I$_c$,  which have been constrained by relic density, to the production of cosmic-ray. Due to $v_\Delta < 10^{-4}$ GeV,  the positron and electron production is dominated by $\delta^{\pm\pm}$ and $\delta^{\pm}$ decays. Since the BRs of triplet decaying to leptons have been  given in Table~\ref{tab:brs},  in order to understand the cross section for producing the  triplet pairs, we 
calculate the normalized cross section in each  scenario and  present the results in Table~\ref{tab:Annihilations}, where the normalisation is defined by $\sigma(SS\to \delta_i \bar\delta_i)/\sum_{i} \sigma(SS\to \delta_i \bar\delta_i)$ with $\delta_i=\delta^{++}, \delta^+,\delta^0, \eta^0$.
\begin{table}[hpbt]
\caption{ Normalized cross section for DM annihilating to triplet-pair  in each scenario.}
\begin{ruledtabular}
\begin{tabular}{ccccc}
channel & $\delta^{++} \delta^{--} $ & $\delta^{+} \delta^{-} $ & $\delta^{0} \delta^{0} $ & $\eta^{0} \eta^{0} $  \\ \hline \hline
 I$_a$ & 4/5 & 1/5 & 0 & 0   \\ \hline
  I$_b$ & 2/6 & 2/6 & 1/6 & 1/6   \\ \hline
    I$_c$ & 0 & 1/5 & 2/5 & 2/5   \\ \hline \hline
  II & 0 & 1/5 & 2/5 & 2/5 
  \end{tabular} 
\end{ruledtabular}
 \label{tab:Annihilations}
\end{table}
\begin{figure}[hpbt] 
\begin{minipage}{0.45 \hsize}
\begin{center}
\includegraphics[width=70mm]{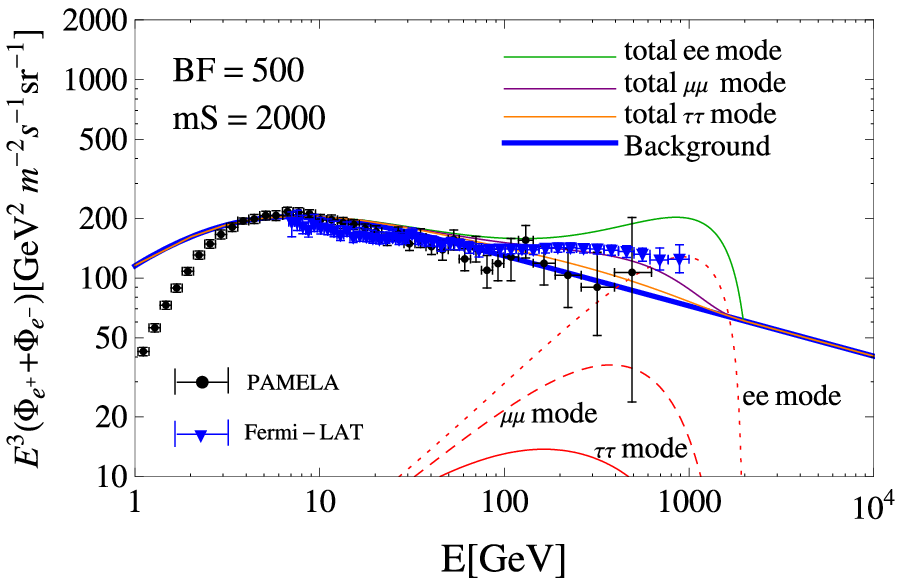} 
\end{center}
\end{minipage}
\begin{minipage}{0.45 \hsize}
\begin{center}
\includegraphics[width=70mm]{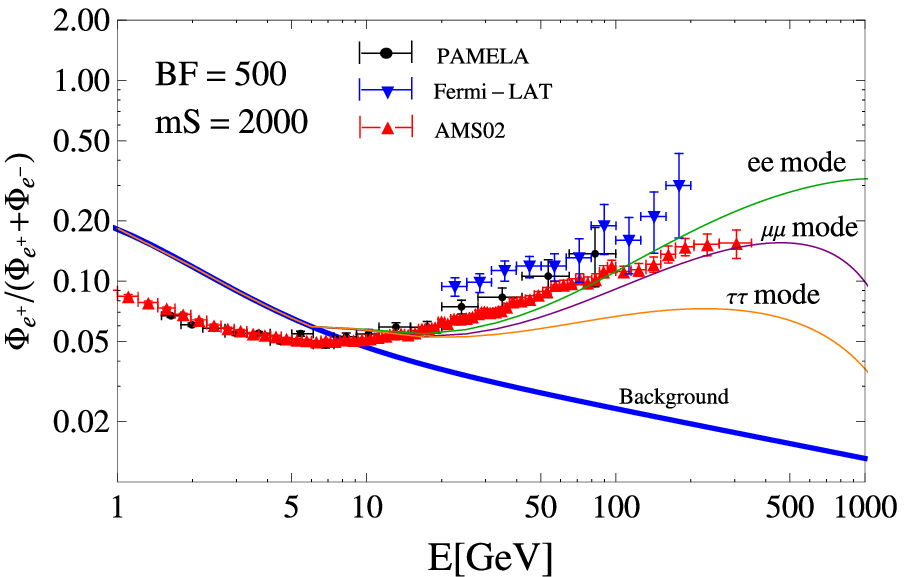}
\end{center}
\end{minipage} 
\caption{ The electron+positron spectra (left) and positron fraction (right) from $\delta^{\pm \pm} \rightarrow \ell^\pm \ell^\pm$ decays, where
we adopt scenario-I$_a$ and NO for neutrino masses and  take $m_S=2000$ GeV and boost factor $BF=500$.  The associated experiments data are  PAMELA~\cite{Adriani:2011xv, Adriani:2008zr} and Fermi-LAT~\cite{Ackermann:2010ij,FermiLAT:2011ab} and AMS02~\cite{Aguilar:2013qda}.
\label{fig:SingleMode}}
\end{figure}

 Combining the constraints of free parameters and  the mass spectra of neutrinos discussed before, we now study the numerical analysis for cosmic-ray positron and electron spectra. First we focus on the positron/electron production via doubly charged scalar $\delta^{++}$. The channel is interesting because not only  it has a larger normalized cross section shown in Table~\ref{tab:Annihilations},  but also there are six different modes to generate positrons/electrons, such as $\delta^{++}\to (e^+ e^+, e^+ \mu^+, e^+ \tau^+, \mu^+ \mu^+, \mu^+ \tau^+,\tau^+ \tau^+)$, where $\mu^+$ and $\tau^+$  then continue decaying to positron by the electroweak interactions in the SM. For demonstrating the behavior of multiple decay chains, we show  $\Phi_{e^+} + \Phi_{e^-}$ and $\Phi_{e^+}/(\Phi_{e^+} + \Phi_{e^-})$ for the modes $\delta^{\pm\pm}\to \ell^\pm \ell^\pm$ with $\ell=e, \mu, \tau$ in Fig.~\ref{fig:SingleMode}, where  for illustration, we choose the scheme I$_a$, NO for neutrino masses, $m_S=2000$ GeV and $BF=500$. 
For comparisons, we also show the data, measured by PAMELA~\cite{Adriani:2011xv} and Fermi-LAT~\cite{Ackermann:2010ij} for $\Phi_{e^-} + \Phi_{e^+}$  and by PAMELA~\cite{Adriani:2008zr}, Fermi-LAT~\cite{FermiLAT:2011ab} and AMS-02~\cite{Aguilar:2013qda} for $\Phi_{e^+}/(\Phi_{e^+} + \Phi_{e^-})$, in the figure. By the results, we clearly see that the curve for positron+electron spectrum from $\mu \mu$ mode is  flatter than  that from $ee$ mode, 
 and the spectrum from $\tau\tau$ mode is more flat and just  slightly over the background.
For $\Phi_{e^+}/(\Phi_{e^+} + \Phi_{e^-})$, $\tau\tau$ mode gives too small contribution to fit the measurements while $ee$ and $\mu\mu$ are much close to the experimental data in the measured region.

%
%
In the following we discuss the fluxes of  cosmic-ray positrons and electrons that  are from all possible sources.
First,  in Table~\ref{tab:BoostFactor} we present the required BFs for fitting the excess measured by PAMELA, Fermi-LAT and AMS-02. We see that the BFs for scenario-I$_c$ and -II have been over the upper bounds ($BF_{\rm max}$)  that are obtained from antiproton measurement. In scenario-I$_{a}$ and -I$_b$, for satisfying the bound of BF,  the  mass of DM should be heavier than 1000 GeV. The values in brackets  in the table denote the required BFs  for neutrino masses with NO, IO, QDI and QDII by turns. It is found that the required BFs in scenarios I$_a$ and I$_b$ are close to each other.
\begin{table}[b] 
\caption{ Required BF in each scenario associated with $m_S$ for explaining positron/electron excess,
where $BF > BF_{\rm max}$ indicates  the necessary BF over the upper limit of antiproton flux.
The values in brackets stand for the required BFs for neutrino masses with NO, IO, QDI and QDII by turns.}
\begin{ruledtabular}
\begin{tabular}{ccccc} 
 $m_S$ & $1000$ GeV & $2000$ GeV & $3000$ GeV & $4000$ GeV  \\ \hline 
 I$_a$  & $BF > BF_{\rm max}$ & $ (500, 500, 500, 500)$ & $ (1400,1200, 900, 1100)$ & $ (2000, 1900, 1500, 1900)$   \\ \hline
  I$_b$  & $BF > BF_{\rm max}$  & $(500, 500, 500, 500)$ & $ (1400, 1200, 900, 1200)$ & $ (2000, 1900, 1500, 1800)$   \\ \hline
  I$_c$ & $BF > BF_{\rm max}$ & $BF > BF_{\rm max}$ & $BF > BF_{\rm max}$ & $BF > BF_{\rm max}$   \\ \hline \hline
  II & $BF > BF_{\rm max}$ & $BF > BF_{\rm max}$ & $BF > BF_{\rm max}$ & $BF > BF_{\rm max}$  \\ \hline
\end{tabular} 
\end{ruledtabular}
 \label{tab:BoostFactor}
\end{table}
\begin{figure}[t] 
\begin{minipage}{0.45 \hsize}
\begin{center}
\includegraphics[width=70mm]{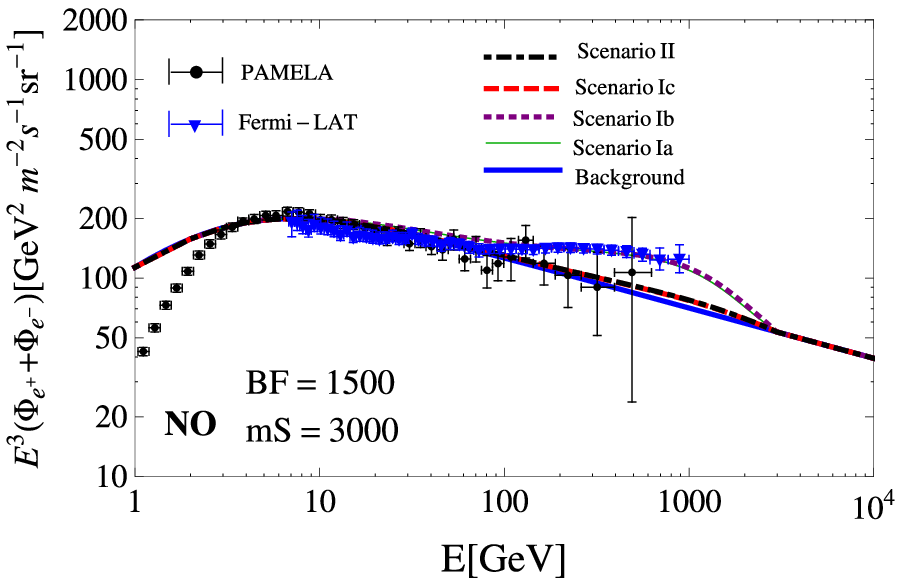} 
\end{center}
\end{minipage}
\begin{minipage}{0.45 \hsize}
\begin{center}
\includegraphics[width=70mm]{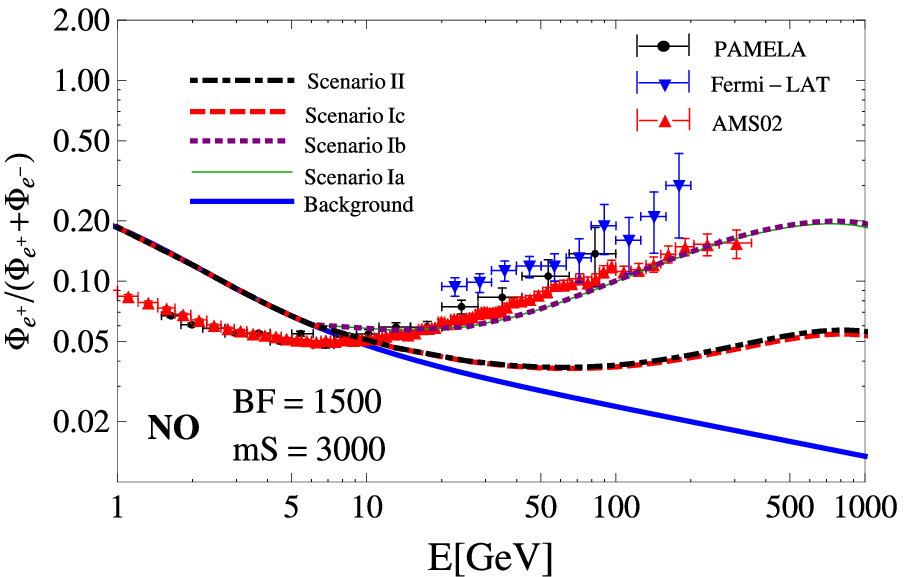}
\end{center}
\end{minipage} 
\caption{ The positron+electron spectrum (left) and positron fraction (right) for normal ordered neutrino masses, where we have used
$m_S=3000$ GeV and boost factor $BF=1500$. The ( solid, dotted, dashed, dot-dashed) line corresponds to scenario (I$_a$, I$_b$, I$_c$, II). The thick solid line is the cosmic-ray background. For left panel, we quote the data of  PAMELA~\cite{Adriani:2011xv} and Fermi-LAT~\cite{Ackermann:2010ij}. For right panel, we quote the data of PAMELA~\cite{Adriani:2008zr}, Fermi-LAT~\cite{FermiLAT:2011ab} and AMS02~\cite{Aguilar:2013qda}. 
\label{fig:positronBP1}}
\end{figure}
\begin{figure}[t] 
\begin{minipage}{0.45 \hsize}
\begin{center}
\includegraphics[width=70mm]{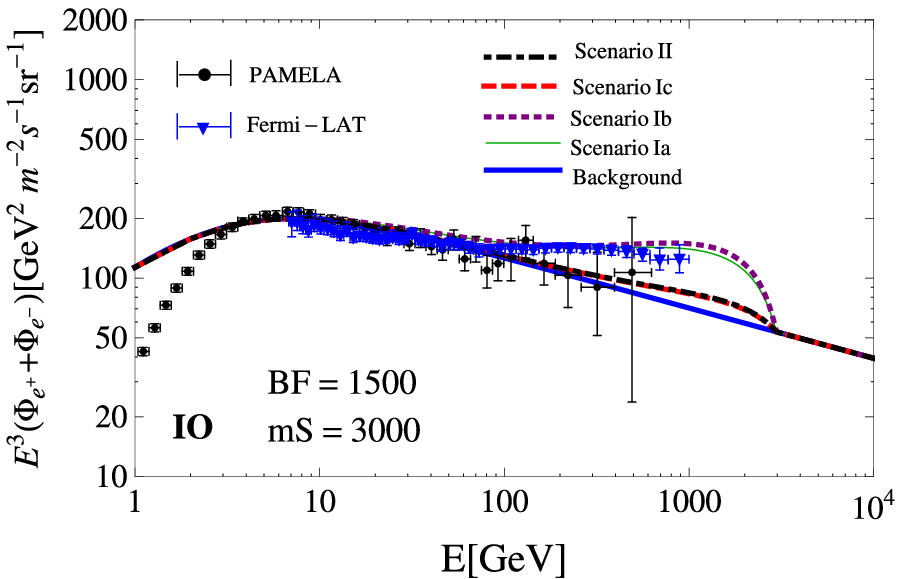} 
\end{center}
\end{minipage}
\begin{minipage}{0.45 \hsize}
\begin{center}
\includegraphics[width=70mm]{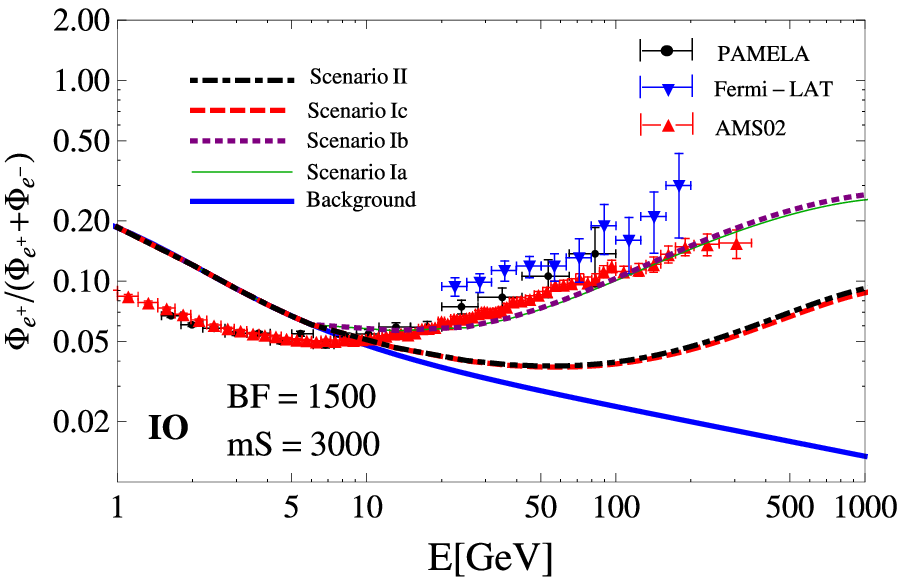}
\end{center}
\end{minipage} 
\caption{ The legend is the same as Fig.~\ref{fig:positronBP1} but for inverted neutrino masses.}
\label{fig:positronBP2}
\end{figure}

\begin{figure}[t] 
\begin{minipage}{0.45 \hsize}
\begin{center}
\includegraphics[width=70mm]{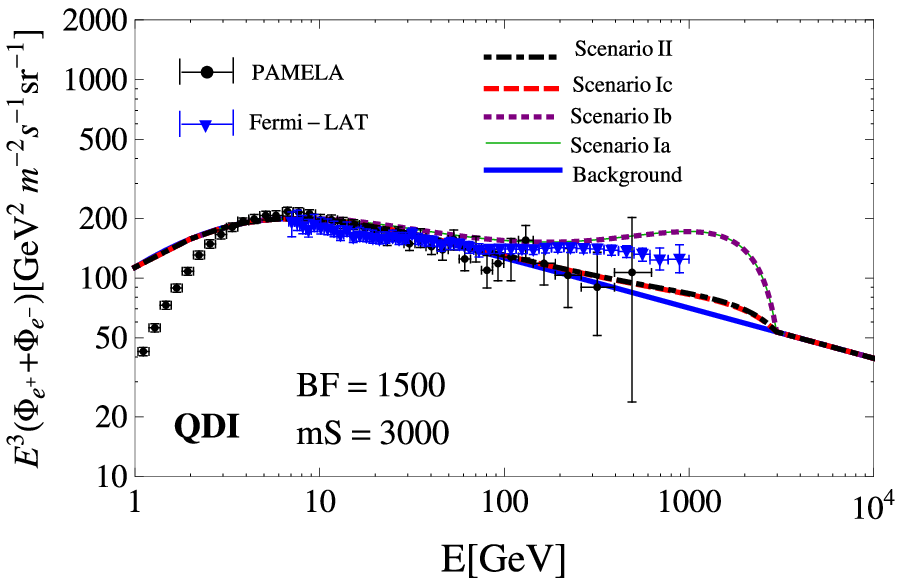} 
\end{center}
\end{minipage}
\begin{minipage}{0.45 \hsize}
\begin{center}
\includegraphics[width=70mm]{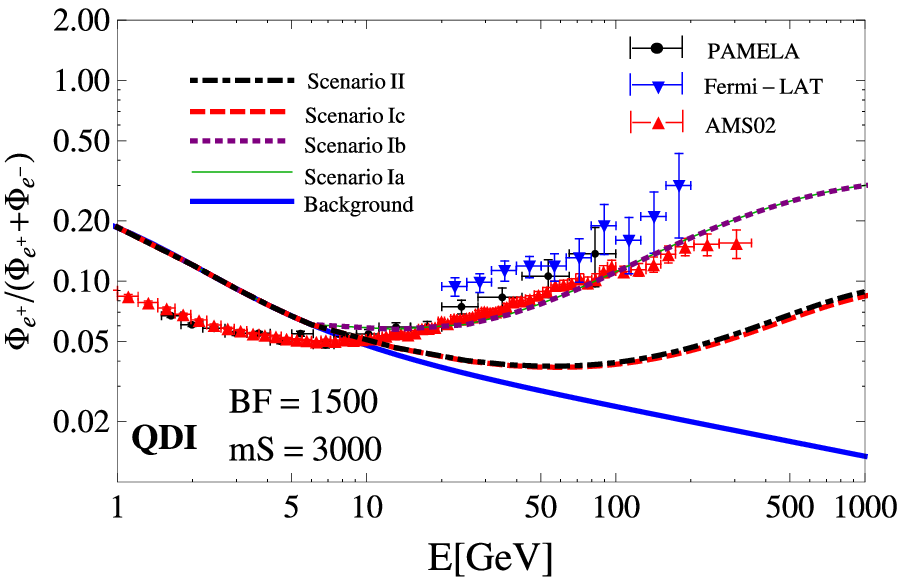}
\end{center}
\end{minipage} 
\caption{ The legend is the same as Fig.~\ref{fig:positronBP1} but for QDI.}
\label{fig:positronBP3}
\end{figure}

\begin{figure}[t] 
\begin{minipage}{0.45 \hsize}
\begin{center}
\includegraphics[width=70mm]{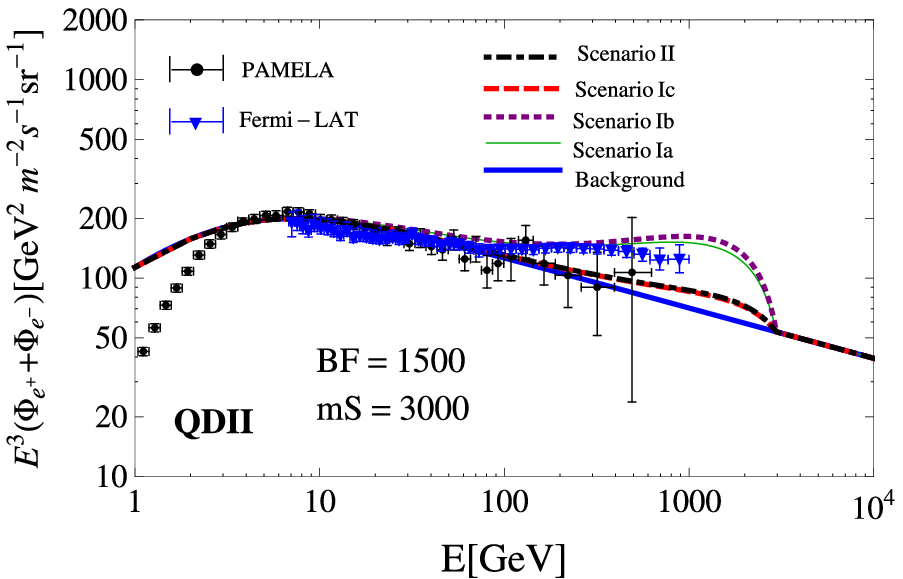} 
\end{center}
\end{minipage}
\begin{minipage}{0.45 \hsize}
\begin{center}
\includegraphics[width=70mm]{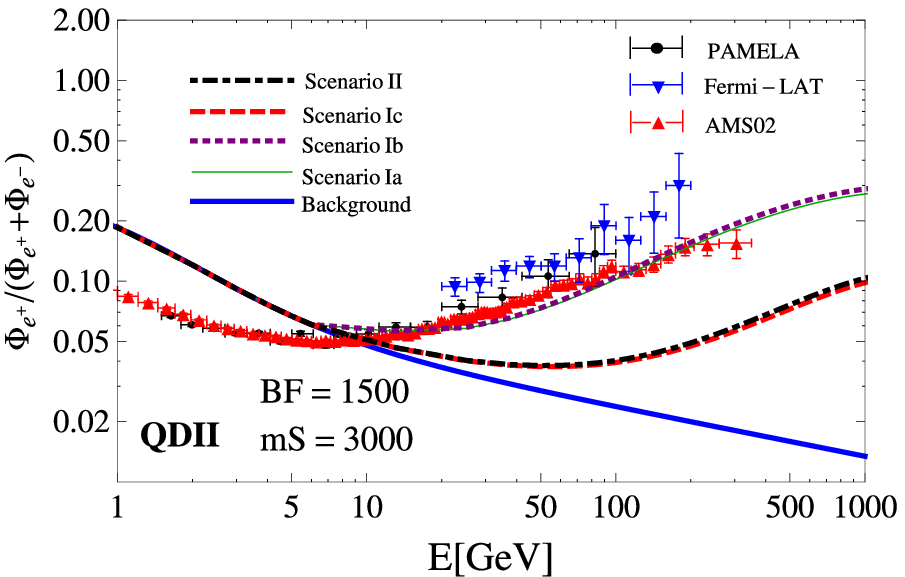}
\end{center}
\end{minipage} 
\caption{ The legend is the same as Fig.~\ref{fig:positronBP1} but for QDII.}
\label{fig:positronBP4}
\end{figure}

\begin{figure}[t] 
\begin{minipage}{0.45 \hsize}
\begin{center}
\includegraphics[width=70mm]{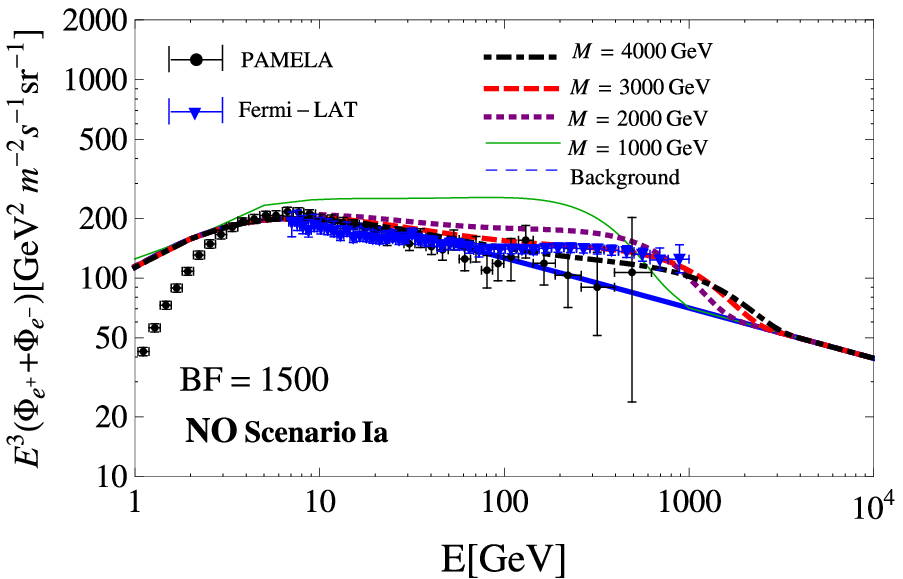} 
\end{center}
\end{minipage}
\begin{minipage}{0.45 \hsize}
\begin{center}
\includegraphics[width=70mm]{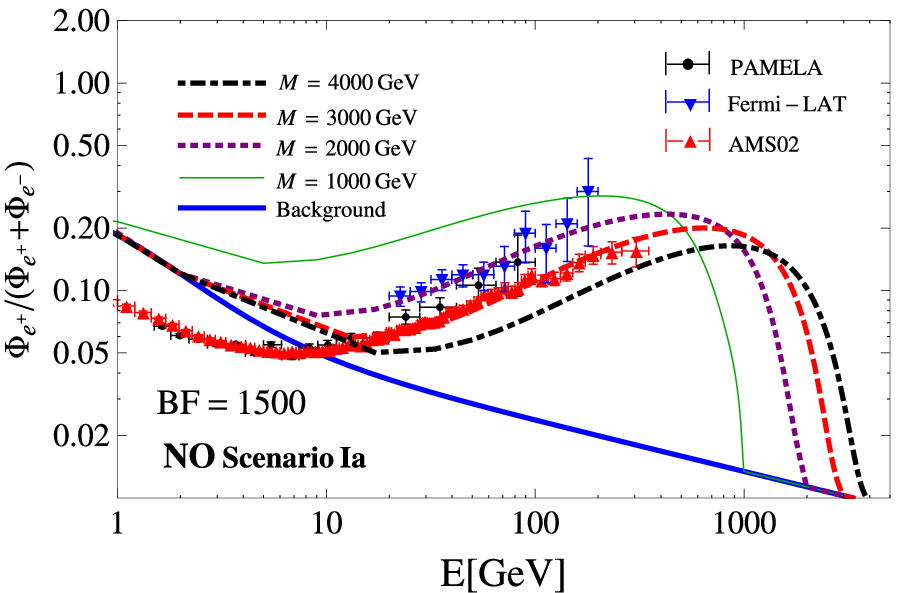}
\end{center}
\end{minipage} 
\caption{ The positron+electron spectrum (left) and positron fraction (right) for normal ordered neutrino masses, where we have used
$m_S=1000$, $2000$, $3000$ and $4000$ GeV and boost factor $BF=1500$.  The experimental data are the same as those in Figs.~\ref{fig:positronBP1}-\ref{fig:positronBP4}.}
\label{fig:positronMD}
\end{figure}

We calculate $E^3(\Phi_{e^+}+\Phi_{e^-})$ and $\Phi_{e^+}/(\Phi_{e^+} + \Phi_{e^-})$  and  show the results as a function of energy in Fig.~\ref{fig:positronBP1}, \ref{fig:positronBP2}, \ref{fig:positronBP3} and \ref{fig:positronBP4} for neutrino masses with NO, IO, QDI and QDII, respectively, where we adopt  $m_S=3000$ GeV and $BF=1500$, the solid, dotted, dashed and dash-dotted lines in turn denote the scenario-I$_a$, -I$_{b}$, -I$_c$ and -II.  The thick solid line is the cosmic-ray background in Eq.~(\ref{eq:bkgrd}). We see that 
the contributions of scenario-I$_c$ and scenario-II are much smaller than the data. According to the results in Figs.~\ref{fig:positronBP1}-\ref{fig:positronBP4}, we conclude that different neutrino mass spectrum could cause  slight difference in positron/electron flux. Nevertheless, the normal ordered mass spectrum has a better matching with current data. 
 We note that positron/electron flux  in higher energy region  tends to be larger when $BR(\Delta \to \ell_e \ell_e)$ is larger.
In Fig.~\ref{fig:positronMD}, we also show $E^3(\Phi_{e^+}+\Phi_{e^-})$ and $\Phi_{e^+}/(\Phi_{e^+} + \Phi_{e^-})$ for $m_S = 1000$, $2000$, $3000$ and $4000$ GeV in scenario-I$_a$ with NO and $BF=1500$.  
We can see that the end point of positron fraction excess corresponds to the mass of DM.
Thus, the measurement of the positron fraction in higher energy region is important to test the model  and tell us the mass of DM.


\subsection{Cosmic-ray neutrinos}

\begin{figure}[t] 
\begin{center}
\includegraphics[width=3.8 in]{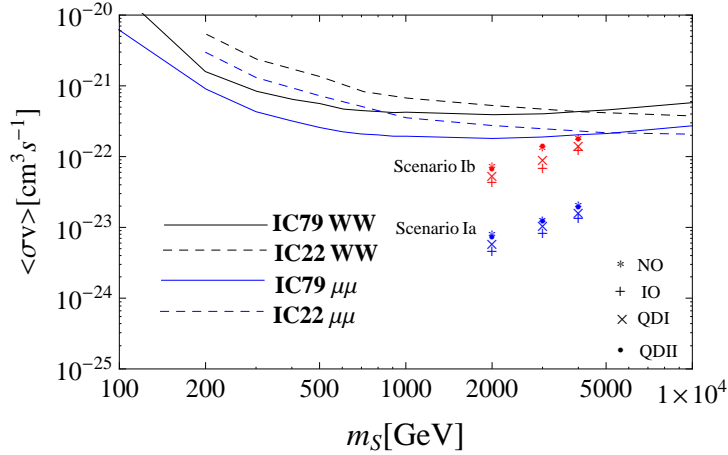} 
\end{center}
\caption{ Velocity-average cross section, $\langle \sigma v \rangle$, annihilating into triplet pairs  for neutrino production in scenario-I$_a$ and I$_b$ with different neutrino mass spectra.  The solid and dashed lines stand for the upper limit of IceCube-22~\cite{Abbasi:2011eq} and IceCube-79~\cite{Aartsen:2013mla} by assuming DM annihilating into $W^+W^-$ and $\mu^+\mu^-$. 
\label{fig:NeutrinoFlux}}
\end{figure}

 As known that the necessary BF in scenario-I$_{c}$ and -II for explaining the positron excess has been excluded by the data of antiproton spectrum, in the following we focus on scenario-I$_{a,b}$ and -III for cosmic-ray neutrinos  induced by  DM annihilation of 
 galactic halo.
For scenario-I$_{a,b}$, the cosmic-ray neutrinos are arisen  from the decays of $\delta^\pm$, $\delta^0$ and $\eta^0$. As known that the original neutrino species from DM annihilation can not be distinguished by experiments, we sum over all possible neutrino final states. It is expected that the results are independent of  the Yukawa couplings in $\Delta\to \nu_i \nu_j$  decays, i.e. independence of neutrino mass spectrum. However, 
the involved parameters for triplet production  are the same as those in the study of  positron flux, where
we have required different BFs in different neutrino mass spectra. Due to the reason, the neutrino production rates in our calculations still depend on the neutrino mass spectra. 
Hence, with the values of BF in Table~\ref{tab:BoostFactor} and with the same values of free parameters  for fitting positron/electron flux, we compute the velocity-averaged cross section   paired triplet production for each scenario and present the results in
Fig.~\ref{fig:NeutrinoFlux}, where  we have used the sum defined by
\begin{equation}
\label{CS_neutrino}
\langle \sigma v\rangle= \langle \sigma v \rangle_{SS \rightarrow \delta^+ \delta^- }+ 2 \langle \sigma v \rangle_{SS \rightarrow \delta^0 \delta^0} + 2 \langle \sigma v \rangle_{SS \rightarrow \eta^0 \eta^0}\,.
\end{equation}
The factor 2 in second and third terms is due to doubled neutrino flux from $\delta^0(\eta^0)$ decays.
 In the figure, we also show the upper limit  of IceCube neutrino flux data  for galactic halo, which are indicated from  $W^+W^-$ and $\mu^+ \mu^-$ pairs \cite{Abbasi:2011eq, Aartsen:2013mla}. 
We clearly see that although both results of I$_{a}$ and I$_b$ are lower than the upper bound of data, the scenario-I$_a$ is much close to the bound.  The IceCube measurement will give a further limit on our parameters when more observational data are analysed. 

\begin{figure}[hpbt] 
\begin{minipage}{0.45 \hsize}
\begin{center}
\includegraphics[width=70mm]{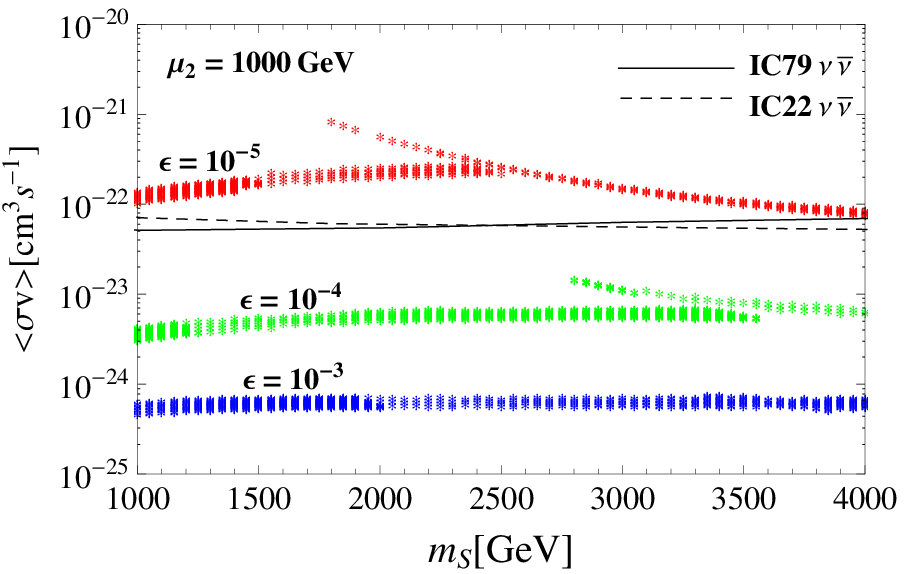} 
\end{center}
\end{minipage}
\begin{minipage}{0.45 \hsize}
\begin{center}
\includegraphics[width=70mm]{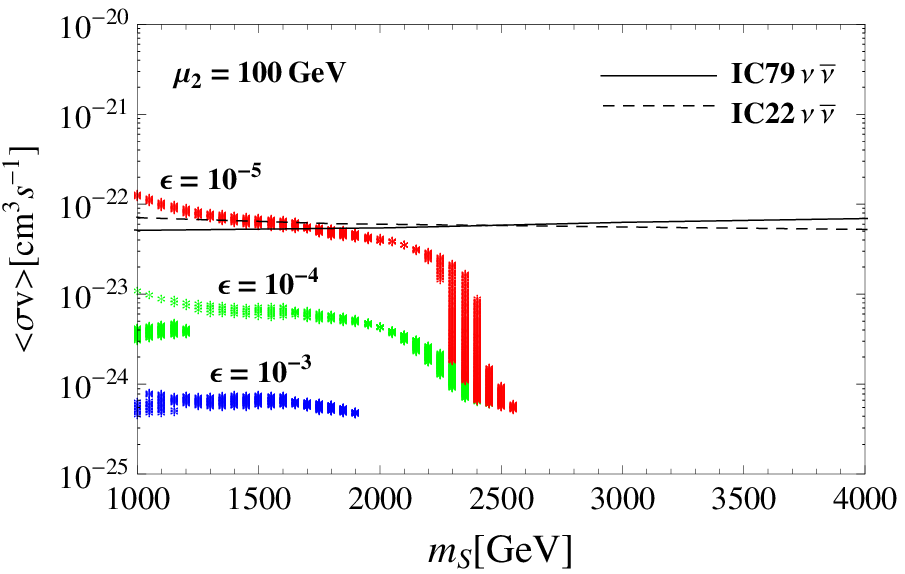}
\end{center}
\end{minipage} 
\caption{ $\langle \sigma v \rangle$ for $SS\to \nu \nu ( \bar\nu \bar\nu)$ as a function of $m_S$ with selected values of $\epsilon$. The left(right) panel is for $\mu_2 = 1000(100)$ GeV. The solid and dashed line stands for the  IceCube upper limit with 79-string~\cite{Aartsen:2013mla} and 22-string~\cite{Abbasi:2011eq}, respectively.}
\label{fig:NeutrinoFlux2}
\end{figure}

For scenario-III, it is known that  when we study the relic density, a Breit-Wigner enhancement appears at $m_\Delta \approx 2 m_S$. It is interesting to investigate the scenario for cosmic-ray neutrinos when the same enhancement  occurs. Unlike the cases in I$_a$ and I$_b$ in which  triplet particles are on-shell, the intermediate state $\delta^0$ in scenario-III could be off-shell, i.e.  the dependence of $h_{\ell' \ell}$ cannot be removed. Additionally, due to the resonant effect which leads to $\langle \sigma v\rangle \propto 1/v^4$ for small width of mediating particle, 
 we find that a large neutrino flux is obtained without BF.
Like the case in relic density, we still use $h_{ee}$ as the free parameter for $h_{\ell' \ell}$ and its constraint  could refer to the Fig.~\ref{fig:allowed2}. 
%
By taking $m_\Delta = 2 m_S (1 - \epsilon)$ and with the  values of parameters constrained by the DM relic density,  $\langle \sigma v \rangle$ for $SS\to \nu\nu (\bar\nu \bar\nu)$ as a function of $m_S$ with several values of $\epsilon$ is displayed in Fig.~\ref{fig:NeutrinoFlux2}, where
the left (right) panel is for $\mu_2 = 1000(100)$ GeV.  The solid and dashed line denotes the IceCube upper limit with 79-string \cite{Aartsen:2013mla} and 22-string \cite{Abbasi:2011eq}, respectively.
%
We note that for $\mu_2 =100$ GeV, the region $m_S \gtrsim 2600$ GeV can not explain the relic density even if there is a Breit-Wigner enhancement.
By the figure, we see that the current IceCube data could limit the value of $\epsilon$. 
We expect that with more data analysis, IceCube could further limit the free parameters of scenario-III. \\


\section{Summary}

For explaining the measured positron excess by the DM annihilation, we have studied  the extension of the SM by adding  an odd IHD $\Phi$ and an even Higgs triplet $\Delta$. 
The LOP of $\Phi$ can be a WIMP DM candidate. Due  to the unbroken $Z_2$-parity, the DM candidate is stable. 
We take the CP-even component $S$ as the DM candidate. The neutrinos become massive
%
through the type-II seesaw mechanism.

In order to suppress the effects from IHD model and emerge the triplet contributions, we have set $\lambda_L=0$, $m_A - m_S=  1$ GeV and $m_{H^\pm}=m_A$. Even though, the antiproton spectrum is dominated by triple interactions $SH^\pm W^\mp$ and $SAZ$ and quadratic interactions $SS(W^\pm W^\mp,ZZ)$ which appear in IHD model. With the measurement of antiproton spectrum, we study the correlation between the upper bound of BF and $m_S$. 

 In terms of the Feynman diagrams, three scenarios are involved in our analysis. In scenario-I, we further  use three schemes to describe the parameters $\chi_A$ and $\chi_B$.  In our model, the excess of positrons/electrons is mainly arisen from the triplet decays. Since the neutrino mass spectrum is still uncertain, we also study the influence of neutrino mass in the cases of NO, IO and QD. From Figs.~\ref{fig:positronBP1}-\ref{fig:positronBP4}, we see that scenario-I$_a$ and -I$_{b}$ have similar contributions. Moreover, the normal ordered mass spectrum could fit well to the excess of positrons/electron measured by PAMELA, Fermi-LAT and AMS-02. 
 
 Although we have not observed the excess of comic-ray neutrinos, however if the source of  excess of positrons/electrons is from triplet decays, the same effects will also increase the abundance of cosmic-ray neutrinos.  We find that the quasi-resonance effects at $m_\Delta \approx 2m_S$ could occur in scenario-III so that 
 large neutrino flux can be obtained without BF.
 We calculate $\langle \sigma v \rangle $ for cosmic-ray neutrinos and realize that our results in some parameter region are close to the recent IceCube data  for neutrino flux from galactic halo.
 Hence, our model could be tested if more data for cosmic-ray neutrinos are observed.  \\

\noindent{\bf Acknowledgments}

 We thank Dr.  A. Pukhov for providing the revised code of micrOMEGAs. This work is supported by the National Science Council of
R.O.C. under Grant \#: NSC-100-2112-M-006-014-MY3 (CHC) and NSC-102-2811-M-006-035 (TN). We also thank the National Center for Theoretical Sciences (NCTS) for supporting the useful facilities.


\end{document}